\documentclass[conference]{IEEEtran}
\IEEEoverridecommandlockouts
\usepackage{cite}
\usepackage{amsmath,amssymb,amsfonts}
\usepackage{algorithmic}
\usepackage{graphicx}
\usepackage{textcomp}
\usepackage{xcolor}
\usepackage{hyperref}
\def\BibTeX{{\rm B\kern-.05em{\sc i\kern-.025em b}\kern-.08em
    T\kern-.1667em\lower.7ex\hbox{E}\kern-.125emX}}

\usepackage{filecontents}

\begin{document}

\title{A System-Level Dynamic Binary Translator using Automatically-Learned Translation Rules}

\author{\IEEEauthorblockN{Jinhu Jiang\textsuperscript{\dag\ddag}, Chaoyi Liang\textsuperscript{\dag}, Rongchao Dong\textsuperscript{\dag}, Zhaohui Yang\textsuperscript{\dag},\\
Zhongjun Zhou\textsuperscript{\dag}, Wenwen Wang\textsuperscript{\#}, Pen-Chung Yew\textsuperscript{\S}, Weihua Zhang\textsuperscript{\dag}}
\IEEEauthorblockA{\textit{School of Computer Science, Fudan University, Shanghai, China\textsuperscript{\dag}} \\
\textit{Institute of Bigdata, Fudan University, Shanghai, China\textsuperscript{\ddag}}\\
\textit{
%Department of Computer Science,
School of Computing, University of Georgia, Athens, GA, USA\textsuperscript{\#}}\\
\textit{Department of Computer Science and Engineering, University of Minnesota, Minneapolis, MN, USA\textsuperscript{\S}}\\
\{jiangjinhu, 18302010035, 20212010113, 18302010057, 19212010010, zhangweihua\}@fudan.edu.cn,\\
wenwen@cs.uga.edu, yew@umn.edu}
}

% \author{\IEEEauthorblockN{1\textsuperscript{st} Given Name Surname}
% \IEEEauthorblockA{\textit{dept. name of organization (of Aff.)} \\
% \textit{name of organization (of Aff.)}\\
% City, Country \\
% email address or ORCID}
% \and
% \IEEEauthorblockN{2\textsuperscript{nd} Given Name Surname}
% \IEEEauthorblockA{\textit{dept. name of organization (of Aff.)} \\
% \textit{name of organization (of Aff.)}\\
% City, Country \\
% email address or ORCID}
% \and
% \IEEEauthorblockN{3\textsuperscript{rd} Given Name Surname}
% \IEEEauthorblockA{\textit{dept. name of organization (of Aff.)} \\
% \textit{name of organization (of Aff.)}\\
% City, Country \\
% email address or ORCID}
% \and
% \IEEEauthorblockN{4\textsuperscript{th} Given Name Surname}
% \IEEEauthorblockA{\textit{dept. name of organization (of Aff.)} \\
% \textit{name of organization (of Aff.)}\\
% City, Country \\
% email address or ORCID}
% \and
% \IEEEauthorblockN{5\textsuperscript{th} Given Name Surname}
% \IEEEauthorblockA{\textit{dept. name of organization (of Aff.)} \\
% \textit{name of organization (of Aff.)}\\
% City, Country \\
% email address or ORCID}
% \and
% \IEEEauthorblockN{6\textsuperscript{th} Given Name Surname}
% \IEEEauthorblockA{\textit{dept. name of organization (of Aff.)} \\
% \textit{name of organization (of Aff.)}\\
% City, Country \\
% email address or ORCID}
% }

\maketitle

% Names
\newcommand{\qemu}{QEMU 6.1}
\newcommand{\spec}{SPEC CINT2006}

% Basic
\newcommand{\coverage}{94.10\%}
\newcommand{\baseSpeedup}{0.95}
\newcommand{\baseSlowdownQemu}{5\%}
\newcommand{\optSpeedup}{1.36}
\newcommand{\appSpeedup}{1.15}

% Host instr per sync
\newcommand{\instrPerSync}{14}

% Host instr of sync per guest instr
\newcommand{\baseSync}{8.36}
\newcommand{\lazySync}{1.79}
\newcommand{\eliminationSync}{1.33}
\newcommand{\schedulingSync}{0.89}

% Opt related
\newcommand{\syncOpt}{48.83\%}
\newcommand{\eliminationSyncOpt} {36.52\%}
\newcommand{\schedulingSyncOpt} {24.61\%}

% Trans related
\newcommand{\baseTransNum}{17.39}
\newcommand{\optTransNum}{15.40}

% Slowdown related
\newcommand{\optSlowdown}{13.83}
\newcommand{\baseSlowdown}{18.62}
\newcommand{\qemuSlowdown}{18.73}

% Editing related
\newcommand{\red}[1]{\textcolor{red}{#1}}
\newcommand{\blue}[1]{\textcolor{blue}{#1}}

% Preserved
\newcommand{\yew}[1]{\color{red}#1}
\newcommand{\wang}[1]{\color{blue}#1}
\newcommand{\zhang}[1]{{\color{orange}{#1}}}

%-------------------------------------------------------------------------------
\begin{abstract}
    %-------------------------------------------------------------------------------
    
    System-level emulators have been used extensively for the design, debugging and evaluation of the system software.
    They work by providing a system-level virtual machine that can support a guest operating system (OS) running on a platform with the same or different native OS using the same or different instruction-set architecture.
    For such a system-level emulation, dynamic binary translation (DBT) is one of the core technologies.
    A recently proposed learning-based approach using automatically-learned translation rules has shown to improve DBT performance significantly with much higher quality translated code. However, it has only been used on \emph{user-level} emulation, not \emph{system-level} emulation.
    %Since the learning-based approach uses host registers to maintain the guest CPU state, it needs to coordinate its CPU state with QEMU's memory CPU state when a guest instruction cannot be translated by its rules.
    %This approach gains high-performance results in the user-level but does not yet support system-level emulation.
    %In this paper, we explore the feasibility of applying this approach to improve system-level emulation, and use QEMU to build a prototype.
    %\blue{However, this approach unexpectedly achieves a \baseSlowdownQemu{} slowdown.}
    
    %\yew{
    In applying this approach directly on QEMU for system-level emulation, we find it actually causes an unexpected performance degradation of 5\% on average.
    By analyzing its main culprits in more detail, 
    we find that the learning-based approach will by default use host registers to maintain the guest CPU states that include condition-code registers (or FLAG registers).
    In cases where QEMU needs to be involved (in which QEMU also needs to use the host registers), maintaining system states in the host registers for the guest, the host and QEMU \emph{during} and \emph{between} the context switches can cause undue overheads, if not handled carefully. 
    Such cases include emulating system-level instructions, address translation and interrupts, which require the use of QEMU's helper functions. 
    %We analyze major challenges in using this approach to support system-level emulation.
    %As the learning-based approach uses host registers to maintain the guest CPU state, it needs to coordinate its guest CPU state with the QEMU's CPU state in memory when a guest instruction is not covered in its translation rules.
    %Also, there exist cases that include emulating system-level instructions and handling address translation and interrupts, which require the use of QEMU's helper functions.
    %In these cases, efficient coordinations for CPU states are required to guarantee correctness.
    %To address those issues, we propose a coordination method to solve these problems.
    %Furthermore, we leverage several optimizations that include lazy-parsing optimization to reduce the overhead of each coordination, and coordination elimination and code scheduling to reduce the coordination frequency.
    To achieve the intended performance improvement through better-quality code generated by the learning-based approach, we propose several optimization techniques that include reducing the overhead incurred in each context switch, the number of needed context switches, and better code scheduling to eliminate context switches.
    %We prototype our design based on the original learning-based approach and support full system emulation successfully.
    Our experimental results show that such optimizations can achieve an average of \optSpeedup X speedup over \qemu~using \spec~and \appSpeedup X on real-world applications in the system emulation mode.
    %}
    \end{abstract}

%-------------------------------------------------------------------------------
\section{Introduction}
%-------------------------------------------------------------------------------
\label{sec:introduction}

System-level emulators are an important tool for designing new system architectures, debugging binary codes and profiling application programs in a full system environment.
A system-level emulator emulates the binary code of a guest operating system (OS) implemented on the guest instruction set architecture (ISA), and  booted on a host with the same or a different ISA running the same or a different OS.
In a system-level emulation, dynamic binary translation (DBT) is one of the core technologies.
This technology has been used in many applications provided by VMware, Valgrind, QEMU\cite{DBLP:conf/usenix/Bellard05}, and Rosetta.
In essence, a DBT system provides the capability of 
dynamically translating a guest binary code to run on a host with a different system environment at runtime. 
%execute guest binary codes on a different host machine by dynamically translating the guest binary code using the host instruction set architecture.

Depending on usage scenarios, a DBT can emulate guest binaries at the \emph{user} level or the \emph{system} level.
When at the \emph{user} level, the DBT only translates the guest binaries and runs the translated binaries directly on the host OS without emulating detailed guest OS operations such as virtual address translation and system calls.
While at the \emph{system} level, the DBT needs to translate the entire guest execution environment that includes all guest OS operations.
It thus provides a complete system-level emulation on top of the host environment.

A general dynamic binary translator, such as QEMU, provides a general framework that includes an intermediate representation (IR) as a common interface between the guest and the host binaries.
Guest binaries are first translated to the QEMU IR, and then from the IR to host binaries in different ISAs, i.e., it is a "many(ISAs)-to-many(ISAs)" binary translation.
%Due to the expression limitation of IR, this DBT approach has a relatively poor translated code quality.
However, due to this two-step translation approach, each guest instruction will be translated into \emph{n} IR instructions and each IR instruction to \emph{m} host instructions, with a total of \emph{n}x\emph{m} host instructions.
For example,~\cite{DBLP:conf/micro/JiangDZSWYZ20} shows that a guest ARM instruction can be translated into 8.18 host x86 instructions on average.
In addition, it requires significant engineering effort to manually create translation rules that translate each guest instruction to the IR, and then translate each IR instruction to the host binaries.

A learning-based DBT approach~\cite{DBLP:conf/micro/JiangDZSWYZ20}\cite{DBLP:conf/usenix/SongWYZZ19}\cite{DBLP:conf/asplos/WangMZY18} was proposed recently to resolve those issues using automatically-learned translation rules for each pair of guest ISA and host ISA, i.e., an "one(ISA)-to-one(ISA)" binary translation approach.
%translate guest binary codes directly to host binary codes using translation rules.
These translation rules can be automatically learned from the optimized guest and host binary codes produced by compilers using the same source code.
It is an 'one-step' translation approach (i.e. without going through IR) based on high-quality translation rules with minimal engineering effort.
However, this approach has only been applied to user-level emulation, and not yet to system-level emulation.
%\blue{
Our recent experimental results show that, although the learning-based approach works well at the user level, it unexpectedly causes a \baseSlowdownQemu{} slowdown after we apply it to a system-level emulation.
%}

%Despite these benefits, the learning-based approach is currently implemented only on user-level DBT and does not yet support system-level DBT.
% We deeply analyze the challenges in supporting the system mode emulation for learning-based method.
%A DBT system needs to maintain the guest CPU state, like guest registers, condition codes/flags that are set implicitly, etc.
%For example, QEMU uses a sequence of memory addresses to maintain the guest CPU state.
%The learning-based approach uses the host registers for CPU state maintenance instead.
%However, since the learning-based approach cannot reach 100\% code coverage, which means it will occasionally switch to QEMU to perform emulation if it fails to translate some instructions.
%Thus, CPU state coordination is necessary when a learning-based approach used in DBT.
% The existing approach works well at the user level, but it is more challenging at a system-level emulation.
% But there are more coordination issues in system emulation, which including system-level instructions, memory address translation, and system interrupts.
% When learning-based approach encounters these scenarios, it has to switch to QEMU, resulting in CPU state inconsistency between learning-based approach and QEMU.
In this paper, we first analyze the new challenges in the system-level emulation.
We find that, when it encounters system-level instructions as well as address translation and interrupts, which are common in a system mode emulation, the learning-based approach needs to switch to QEMU for various system support.
%The existing learning-based approach can encounter several issues during such context switches.
Any DBT system, not just learning-based DBT systems, needs to maintain guest CPU states such as the content of general registers and condition codes/flags registers that are set implicitly. 
For example, QEMU uses a designated memory region to hold and maintain such guest CPU states.
The learning-based approach, on the other hand, uses the \emph{host registers} to hold and maintain guest CPU states by default.
%However, since the learning-based approach cannot reach 100\% code coverage, which means it will occasionally switch to QEMU to perform emulation if it fails to translate some instructions.
%Thus, CPU state coordination is necessary when a learning-based approach used in DBT.
%For example, QEMU uses a sequence of memory addresses to maintain the guest CPU state.
%The learning-based approach uses the host registers for CPU state maintenance instead.
In cases where QEMU needs to be involved (in which QEMU also needs to use the host registers), maintaining system states in the host registers for the guest, the host and QEMU \emph{during} and \emph{between} the context switches can cause undue overheads.
These overheads can offset the benefits derived from the learning-based approach if not coordinated and handled carefully. 

To achieve a better performance, we propose several optimizations to reduce the overhead incurred \emph{during} and \emph{between} the context switches.
%Among them, \emph{lazy parsing} can delay the parsing of the CPU state to reduce the number of needed instructions in each coordination.
%Among them, \emph{coordination overhead reduction} can delay the parsing of the CPU state to reduce the number of needed instructions in each coordination.
One optimization is to delay the parsing of the guest CPU state. It can reduce the number of needed instructions to maintain the guest CPU state during each context switch.
%\emph{Coordination elimination} 
We also identify several common scenarios that can create rapid consecutive context switches and incur a substantial amount of overhead. 
They include (1) consecutive memory access instructions that require emulating the address translation in QEMU, (2) define-before-use translation blocks (TBs), and (3) consecutive context switching between the translated code and QEMU due to system calls, interrupts or instruction sequences not in the translation rules. 
%These scenarios can create rapid consecutive context switches and incur a substantial amount of overheads.
We can consolidate and combine some of those episodes to reduce the number of context switches and their associated overheads.
%then eliminate the redundant coordinations incurred in those scenarios to reduce the required CPU state coordination.
Further optimizations include doing a better code scheduling for guest instructions that define and use CPU states (e.g. condition codes/flags), which can reduce redundant instructions needed to maintain those CPU states.
%Merging the interrupt check function to the nearest memory access instruction.

To study the effectiveness and the performance improvement that the learning-based approach can achieve with those optimization techniques in a system-level DBT, we implemented a prototype on QEMU 6.1.
We use \spec~to evaluate our design.
%Experimental results show that the dynamic coverage of the learning-based approach can reach \coverage ~on average.
Experimental results show that, compared to the QEMU running in a system mode, the learning-based approach without our proposed optimizations has an average of \baseSlowdownQemu{} slowdown.
When all optimizations are applied, \syncOpt~of all operations that are related to maintaining guest CPU states can be eliminated.
It can achieve a performance improvement of \optSpeedup X over the baseline QEMU on average.
%\blue{
In addition, we also use several real-world applications for evaluation.
The results show that the learning-based approach achieves an average of \appSpeedup X speedup.
%}
% The results show that although our learning-based approach works well on CPU-bound workloads.
% For those IO-bound or network workloads, however, the speedup brought by learing-based approach is realtivly lower even less than 10\%.
% This is because the learing-based approach focus on improving code quality, but large time blocked on IO will reduce benefits of learning-based approach.

In summary, this paper makes the following contributions:

\begin{itemize}
	% \item We propose a learning-based, system-level DBT scheme that directly maintains the guest CPU states using the host CPU states without going through the  memory. Moreover, we come up with several optimizations to reduce the overhead of the required CPU state coordinations.
	\item We apply the learning-based approach to system-level DBT, and propose several optimizations to reduce the overhead required to maintain CPU states \emph{during} and \emph{between} the context switches frequently encountered in the system mode.

	\item We implement the proposed learning-based system-level design and its optimizations in a prototype using QEMU. 
 %which applies the learning-based DBT approach to a system-level emulator for the first time.
	
	\item We conduct several experiments to evaluate the learning-based system-level emulator.
	The results show that we can achieve an average of \optSpeedup X speedup over QEMU on \spec{} and \appSpeedup X on real-world applications with all of the proposed optimizations applied.
\end{itemize}

The rest of this paper is organized as follows.
In Section~\ref{sec:background}, we discuss the motivation and challenges in applying the learning-based approach to a system-level DBT.
In Section~\ref{sec:design_opt}, we present the design of the learning-based system-level emulator and propose several optimizations to improve the coordination efficiency.
In Section~\ref{sec:evaluation}, we evaluate our implementation and show some experimental results.
Section~\ref{sec:related} includes some related work. Section~\ref{sec:conclusion} concludes this paper.

%-------------------------------------------------------------------------------
\section{Background \& Motivation}
%-------------------------------------------------------------------------------
\label{sec:background}

In this section, we first give a brief introduction to the learning-based DBT approach,
and then discuss the CPU state maintenance strategy in the learning-based DBT approach.
Finally, we present the challenges of applying the learning-based approach to system-level DBT.

%------------------------------------------------------------------------------------------------------------
%------------------------------------------------------------------------------------------------------------
\subsection{Learning-Based DBT Approach}

There are three parts in the learning-based DBT approach: (1) learning of the translation rules, (2) parameterization of the learned rules, and (3) rules application~\cite{DBLP:conf/micro/JiangDZSWYZ20}\cite{DBLP:conf/usenix/SongWYZZ19}\cite{DBLP:conf/asplos/WangMZY18}.

In the learning phase, it uses an automated learning framework to generate high-quality translation rules. %, as shown in Figure~\ref{fig:learning}.
First, it compiles a source code into executable binaries of both the \emph{guest} and the \emph{host} architectures (e.g., ARM and x86) using popular compilers such as LLVM or GCC with the debug option turned on.
%It typically with high optimization compilation options uses compilers (e.g., LLVM-ARM and LLVM-x86) .
Next, it uses the generated debugging information, such as the line numbers of the source code, to extract the semantically-equivalent code fragments in the two binaries, e.g., the binary sequences that correspond to the same source statement.
These two code fragments form the basis of a translation rule since they are from the same source statement and are supposed to be semantically equivalent.
%These not-verified code mapping relationship can also be called candidate translation rules.
 %Third, it chooses initial mappings for the operands of the candidate translation rules, 
It then uses a symbolic execution tool to perform a formal semantic-equivalence verification of the two code fragments.
After the formal verification, the two code fragments (i.e., the guest and the host binary sequences) finally form a translation rule.
%and the harvested rule is added to the translation rule set.
This process can be iterated automatically using different training source codes to build a more comprehensive and complete translation rule set.
%Due to the automated learning process and the mapping of highly optimized binary instructions, these translation rules have the advantages of low labor costs and high code quality.

% \begin{figure}[!ht]
%     \centering
%     \includegraphics[width=0.477\textwidth]{figures/learning}
%     \caption{Learning translation rules.}
% 	\Description{Learning translation rules.}
%     \label{fig:learning}
% \end{figure}

In the parameterization phase, the translation rules are parameterized to reduce the total number of rules in the rule set and achieve a higher coverage with a smaller training set.
%For example, all of the ALU-related instructions such as \texttt{add}, \texttt{and}, \texttt{or}, etc., can be "parameterized" (or "categorized") into \emph{one} ALU-type instruction. 
Instead of making each of these ALU-type instructions, such as \texttt{add}, \texttt{and}, and \texttt{or}, into a different translation rule, we can lump them together into \emph{one} translation rule for all ALU-type instructions.
In the rule-application phase during the binary translation, we first try to find a matched translation rule in the rule set.
If we cannot find it, it will be switched to QEMU for emulation.
And as mentioned earlier, this context switch will require the guest CPU states to be saved and restored.
%a mixed translation approach is adopted.
%The automatically learned translation rules go first to directly translate guest binaries to host binaries.
%If learning-based approach fails, i.e., rules fail to match the guest instructions, then they will be translated by QEMU.
%\blue{
Such a learning-based approach can significantly improve DBT performance with translation rules learned from the native compilers.
As its focus is on instruction-level translation, it is  orthogonal  to other optimizations such as memory-related optimizations~\cite{DBLP:conf/vee/ChangWHLY14}, parallelism exploitation ~\cite{DBLP:conf/ppopp/WangLCWCZZ11}, and leveraging special host hardware features~\cite{DBLP:conf/vee/DAntrasGGGL17, DBLP:conf/pldi/DAntrasGGL17}.  
%}
%------------------------------------------------------------------------------------------------------------
%------------------------------------------------------------------------------------------------------------
\subsection{CPU State Coordination}

The guest CPU states contain all of the information needed to emulate guest binary codes.
It includes the content of the \emph{general-purpose registers}, status flags in the \emph{condition code register} (CCR) and the \emph{program counter} (PC).
Some DBT systems, such as QEMU, maintain the guest CPU states in the \emph{memory}.
In QEMU, it maps the guest CPU states to a data structure in IR and maintains it in the memory.
%It maintains the data structure in the host registers when it generates host binaries.
When executing the translated host binaries, it loads the guest CPU states from the memory.
After the execution, it stores the latest guest CPU states back in the memory.
Although this strategy is intuitive and easy to implement, it can generate a large number of memory operations and incur a significant overhead.

%Due to the feature of directly translating guest binaries to host binaries, 
As the learning-based approach bypasses the IR and translates the guest binaries directly into the host binaries, it keeps the guest CPU states in the host CPU states as much as possible.
This strategy can reduce the number of memory operations needed to maintain the CPU states.
However, since the learning-based approach cannot achieve a 100\% coverage, 
it will need to switch to QEMU to translate those instructions not covered in its rule set.
The context switching between the translated binaries and QEMU requires additional overhead to maintain correct CPU states.

\begin{figure}[!ht]
    \centering
    \includegraphics[width=0.477\textwidth]{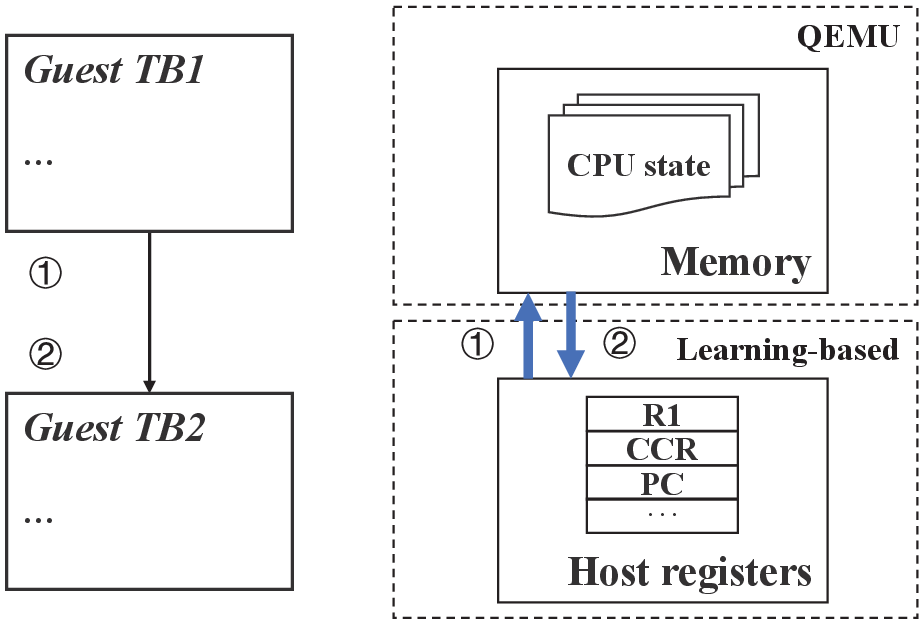}
    \caption{CPU state maintenance and coordination.}
    \label{fig:cpu_state_maintenance}
\end{figure}

%Since the learning-based DBT needs to switch to QEMU when it fails, it needs to coordinate the CPU state to ensure that the learning-based approach can cooperate with QEMU.
%On the one hand, it is used to prevent the native codes that find and translate the next TB from corrupting the CPU state maintained in the host registers by the learning-based approach.
%On the other hand, it is necessary to inform QEMU of the latest CPU state so that QEMU could translate guest instructions with which learning-based approach cannot deal.
We use the example in Figure~\ref{fig:cpu_state_maintenance} to show how it works.
Note that, like most DBT systems, QEMU maintains the translated binaries in a \emph{code cache}. 
%The unit of translated binaries organized in the code cache is a \emph{basic block} of the \emph{guest} binaries, marked as \emph{TB1} and \emph{TB2} in Figure~\ref{fig:cpu_state_maintenance} ("TB" stands for "Thread Block"). 
The unit of translated binaries organized in the code cache is a \emph{basic block} of the \emph{guest} binaries, marked as \emph{TB1} and \emph{TB2} in Figure~\ref{fig:cpu_state_maintenance}.
In this example, after executing the host binary in \emph{TB1} we need to find and translate \emph{TB2}, which will bring it back to QEMU.

At this point, the \emph{guest} CPU state is being maintained in the host CPU registers as shown in the figure.
After the host CPU is switched to QEMU, it will modify the host registers and corrupt the guest CPU state.
Thus, at the end of \emph{TB1}, we need to upload the guest CPU states to the memory locations where QEMU maintains the guest CPU states, as \emph{Path 1} shows.
After QEMU translates \emph{TB2} and places the translated binary in the code cache, it needs to download the guest CPU states from the memory to the host registers where the translated binary in \emph{TB2} maintains the guest CPU states as \emph{Path 2} shows.
% \textcolor{brown}{Assume there is an instruction (marked as \emph{Instr2.}) in \emph{TB2} that cannot be matched by any rule, which needs QEMU to translate it.
% Before switching to QEMU, the learning-based approach needs to coordinate the CPU state to QEMU so that QEMU can correctly translate and execute the instruction.
% After returning from QEMU, the learning-based approach requires to coordinate the CPU state from QEMU to ensure the learning-based approach can possess the latest CPU state.}
% \textcolor{red}{(?????? NOTE: I think this case seems rather redundant. We can remove the brown text. Remember to update Figure 2 accordingly though.???????)}
We call the process of keeping the guest CPU states consistent during such a context switch \emph{"CPU state coordination"}.

%------------------------------------------------------------------------------------------------------------
%------------------------------------------------------------------------------------------------------------
\subsection{Issues and Challenges}
\label{sec:background_issues}

%\blue{
There are several challenges in CPU state coordination when we apply the learning-based approach directly in a system-level DBT.
They can incur a large amount of overhead and  lead to a \baseSlowdownQemu{} slowdown on QEMU as mentioned earlier.
%}
%Specifically, there are several coordination challenges to applying the learning-based approach in system-level DBT.

\textbf{System-level instructions}.
System-level instructions usually perform privileged operations in guest OS, which do not exist in user-level applications.
As a result, the learning-based approach cannot automatically learn system-level instruction rules from user-level applications.
It thus needs to use QEMU to translate the system-level instructions.
QEMU uses a series of helper functions to emulate these instructions.
For example, a privileged ARM instruction such as \texttt{vmsr}\footnote {\texttt{vmsr} transfers the content of an ARM register to its VFP system register} is emulated by a QEMU helper function instead of translating into its corresponding host instruction sequence.
When executing a helper function, it will context switch from the translated binaries in the code cache to QEMU.
If the helper function needs to read or update guest CPU states, the guest CPU states can become inconsistent.
%The coordination design in the user-level DBT does not deal with this inconsistency.

\begin{figure}[!ht]
    \centering
    \includegraphics[width=0.477\textwidth]{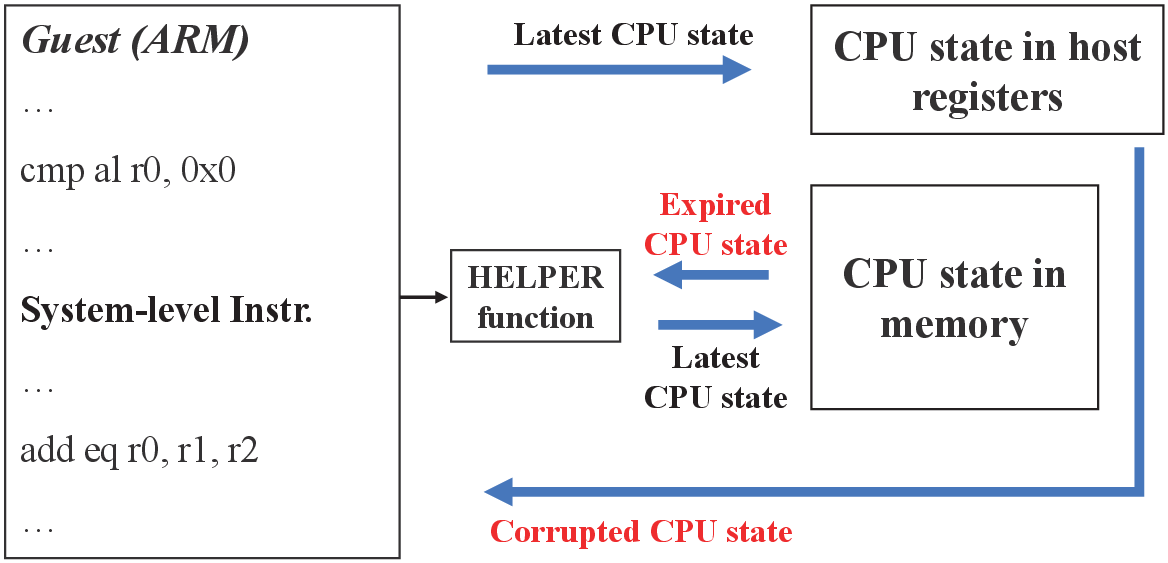}
    \caption{An example of handling a system-level instruction.}
    \label{fig:sys_level_issue}
\end{figure}

Figure~\ref{fig:sys_level_issue} gives such an example. %of this challenge of system-level instructions.
Assume the guest ISA is ARM.
System-level instructions such as \texttt{vmsr} transfer the data between a VFP system register and an ARM register, in which VFP is a vector floating-point system register in an ARM processor. 
Assume that the instructions in the TB are all translated by the translation rules.
When emulating the \texttt{vmsr} instruction, the learning-based approach invokes a QEMU-provided helper function.
The helper function reads the ARM CPU state in the memory maintained by QEMU.
However, this CPU state has expired because an earlier instruction "\texttt{cmp al}" translated by rules will produce a new CPU state, and the learning-based approach maintains the latest guest CPU state in host registers.
On the other side, the helper function for \texttt{vmsr} needs to maintain its own CPU state, which will overwrite the host registers and corrupt the guest CPU state stored there.
%When the translation context switches back from the helper function, 
The following instruction "\texttt{add eq}" after the helper function will have lost the guest CPU state produced by the "\texttt{cmp al}" instruction.
%before the helper function. 
%translated by rules reads the CPU state in host registers.
%However, it is a corrupted CPU state because of the helper function, and QEMU maintains the latest CPU state in memory.
%In both cases above, the CPU state is inconsistent between the learning-based approach and QEMU.

\begin{figure}[!ht]
    \centering
    \includegraphics[width=0.477\textwidth]{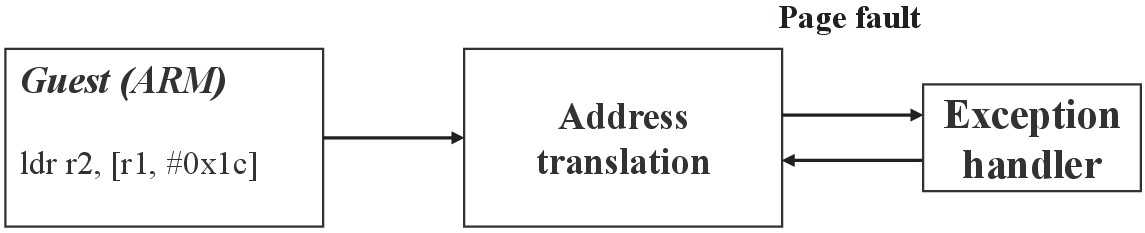}
    \caption{Address translation.}
    \label{fig:addr_trans_issue}
\end{figure}

\textbf{Address translation}.
In a \emph{user-level} DBT, it translates a memory address directly from its \emph{guest virtual address} (GVA) to a \emph{host virtual address} (HVA) by adding a fixed offset.
%It is easy for learning-based approach to use a offset to translate memory access instructions directly.
However, in the \emph{system-level} emulation, the DBT  needs to emulate the guest \emph{memory management unit} (MMU) for potential page faults. In this case, GVA is not necessarily mapped to a pre-determined HVA location.
%(i.e., with a constant offset as in the user level).
Therefore, it needs to be translated through an address translation process as shown in Figure~\ref{fig:addr_trans_issue}. % this paragraph maybe need to re-construction
% In QEMU, the address translation process consists of a fast path and a slow path.
% The fast path converts GVA to HVA by querying the Translation Lookaside Buffer (TLB).
% If the TLB misses, then the slow path is executed.
% It performs a complex address translation process to convert the GVA to the corresponding HVA step by step and caches the GVA-HVA pair of the memory address in the TLB.

The learning-based approach does not support the address translation process.
When encountering a memory access instruction it will context switch to QEMU, and QEMU needs to use the host registers to carry out address translation thus corrupting the guest CPU states.
%Besides, exceptions, e.g., page fault, may occur in the address translation process.
%It also causes the CPU state to be inconsistent between the learning-based approach and QEMU.

% The address translation process brings many coordination challenges.
% As mentioned in figure~\ref{fig:cpu_state_maintenance}, to prevent the CPU state from being corrupted between TBs, at the end of TB, learning-based approach coordinate the CPU state in host registers to memory maintained by QEMU.
% However, the fast path will interrupt the learning-based approach context and execute the TLB-query instructions that may modify the CPU state maintained by learning-based approach if the values of host registers are not coordinated to memory in time.
% Besides, if the previous instruction is translated by rules and defines a new CPU state, then the latest CPU state is maintained by learning-based approach.
% However, the slow path may read the CPU state maintained by QEMU, which is expired and may cause a wrong execution flow.

\begin{figure}[!ht]
    \centering
    \includegraphics[width=0.477\textwidth]{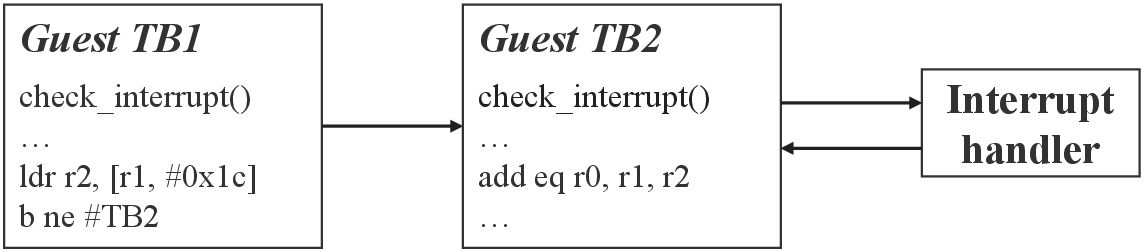}
    \caption{System-level interrupt.}
    \label{fig:interrupt_issue}
\end{figure}

\textbf{Interrupts}.
Figure~\ref{fig:interrupt_issue} shows the interrupt handling mechanism in QEMU.
Interrupts, such as I/O interrupts caused by the keyboard, are caught by the interrupt-check function (i.e., \texttt{check\_interrupt()}) at the beginning of every TB as shown in the figure.
The interrupt check function will invoke the corresponding interrupt handler to deal with the particular interrupt.
QEMU has to translate and execute the interrupt handler provided by the \emph{guest} OS.
If system-level instructions are involved, a context switch from the translated code cache to QEMU is required.
%If interrupts occur in system-level DBT, the translation context will switch from the learning-based approach to QEMU.

%Based on our observation, there is no interrupt in user-level DBT, and the proir learning-based approach does not propose a mechanism for interrupt handling.
%However, in system-level DBT, interrupts will occur and be caught by the interrupt check functions.
%Then it has to translate and execute the interrupt handler provided by the guest OS to handle the caught interrupt.
%Therefore, if interrupts occur in system-level DBT, the translation context will switch from the learning-based approach to QEMU.
%The interrupt handling process also causes CPU state inconsistency between the learning-based approach and QEMU.
%Interrupt handlers break into the learning-based approach context and will corrupt the CPU state in the host registers maintained by learning-based approach.
%Besides, interrupt handlers may use the CPU state maintained by QEMU, which means the interrupt handler may get an expired CPU state if the latest CPU state is defined and maintained by the learning-based approach.

%\textcolor{blue}{
To estimate the overhead of such context switches, 
%To find out the frequency of occurrences in each category mentioned above that require guest CPU state coordination between the code cache and QEMU, 
%i.e., system-level instructions, memory access instructions and interrupt checks, 
we collect at runtime the \emph{dynamic numbers} of 
%\emph{system-level instructions}, \emph{memory access instructions}, \emph{interrupt-check functions}, and \emph{guest instructions} 
system-level instructions, memory access instructions, interrupt-check functions, and the total number of guest instructions executed in each application of~\spec.
We then calculate the occurrence frequency of each category per guest instruction, e.g., \emph{\# of system-level instructions / \# of guest instructions}.
The data are shown in Table~\ref{tab:issue_percentage}.
%} 
%\textcolor{red}{
%each coordination issue mentioned above per guest instruction.
%The statistics for each of the three issues, from high to low, is

\begin{table}[!ht]
	\centering
	\small
	\caption{Distribution of the three categories that require guest CPU state coordination in \spec.}
	\label{tab:issue_percentage}
	\scalebox{0.85}{
	\begin{tabular}{|c|c|c|c|c|}
		\hline
		\textbf{Benchmark} & \textbf{System-level instr.} & \textbf{Memory instr.} & \textbf{Interrupt check} \\ \hline
		perlbench   & 0.28\% & 36.94\% & 19.64\% \\ \hline
		bzip2       & 0.28\% & 40.03\% & 14.24\% \\ \hline
		gcc         & 2.48\% & 29.90\% & 20.11\% \\ \hline
		mcf         & 0.45\% & 41.19\% & 20.53\% \\ \hline
		gobmk       & 0.25\% & 30.58\% & 17.53\% \\ \hline
		hmmer       & 0.09\% & 47.98\% & 5.18\%  \\ \hline
		sjeng       & 0.17\% & 33.86\% & 17.84\% \\ \hline
		libquantum  & 0.09\% & 23.36\% & 9.19\%  \\ \hline
		h264ref     & 0.13\% & 55.21\% & 9.15\%  \\ \hline
		omnetpp     & 0.24\% & 22.54\% & 22.02\% \\ \hline
		astar       & 0.24\% & 31.42\% & 15.92\% \\ \hline
		xalancbmk   & 0.34\% & 23.81\% & 25.94\% \\ \hline
		GEOMEAN     & 0.25\% & 33.46\% & 15.12\% \\ \hline
	\end{tabular}}
\end{table}

%Using geometric means, it shows 33.46\% of the guest instructions are memory access instructions, 15.12\% are interrupt checks, and only around 0.25\% are system-level instructions.
From the statistics, we can see that a context switch is required roughly every two guest instructions on average for QEMU at the system level.
Most of the context switches are for memory access instructions (33.46\%) and interrupt checks (15.12\%). 
Only a small percentage is for system-level instructions (0.25\%) in SPEC CINT2006. 
%Furthermore, the required CPU state coordination and its incurred overhead appear quite substantial.
%}
%little coordination is needed to solve the challenges of system-level instructions.
%In contrast, we require much more coordination to solve the challenges of interrupt checks and memory access instructions.}

%-------------------------------------------------------------------------------
\section{Design \& Optimizations}
%-------------------------------------------------------------------------------
\label{sec:design_opt}

As mentioned earlier, the learning-based approach includes three phases: rule learning, parameterization, and rule application.
CPU state coordination does not affect the rule learning and the parameterization phases.
Thus, they will be the same as in the user-level DBT.
However, in the rule-application phase, we propose an enhanced design with the optimizations mentioned earlier to  reduce the overall context switch overheads.
%make the switch between the code cache and QEMU work more seamlessly at the system level with reduced overheads.

%To address the issues incurred in the required CPU state coordination using the learning-based approach at the system level, we propose an enhanced design and three \blue{architecture-independent} optimization strategies to cut down the overhead.
%To address the challenge of reducing the overhead incurred in the required CPU state coordination using the learning-based approach at the system level, we propose a design and three optimization strategies to cut down the overhead.
%deal with the coordination issues via inserting coordination codes.
% After an in-depth analysis we discover many redundant coordination operations.
% Therefore, we carry out a series of optimizations to improve the code quality.
%In this section, we demonstrate our coordination design and elaborate on how to reduce the overhead of the coordination in the base design using three optimizations.

%------------------------------------------------------------------------------------------------------------
%------------------------------------------------------------------------------------------------------------
\subsection{Basic Guest CPU State Coordination}
\label{sec:basic_design}

%To do that, we utilize an additional coordination mechanism in the third phase and make the system work successfully.

There are two types of guest CPU state coordination.
% As \emph{Path 1} and \emph{Path 2} shown in figure~\ref{fig:cpu_state_maintenance}.
One is the coordination needed when it switches from the code cache to QEMU shown as \emph{Path 1} in Figure~\ref{fig:cpu_state_maintenance}, we call it \emph{sync-save} (as viewed from the perspective of the code cache).
The other is to switch from QEMU back to the code cache shown as \emph{Path 2} in Figure~\ref{fig:cpu_state_maintenance}, 
called \emph{sync-restore}.
% Firstly, the possession of the latest CPU state will be checked.
The needed operations are determined by the context that has the latest guest CPU state before it is switched to the other context.
%If the learning-based approach holds the latest CPU states, we require to pass the values of the corresponding host registers to the memory location where QEMU maintains them, which can be called \emph{sync-save} (from the perspective of the learning-based approach).
%Otherwise, it needs to do the reserve, which is named \emph{sync-restore}.

\begin{figure}[!ht]
    \centering
    \includegraphics[width=0.477\textwidth]{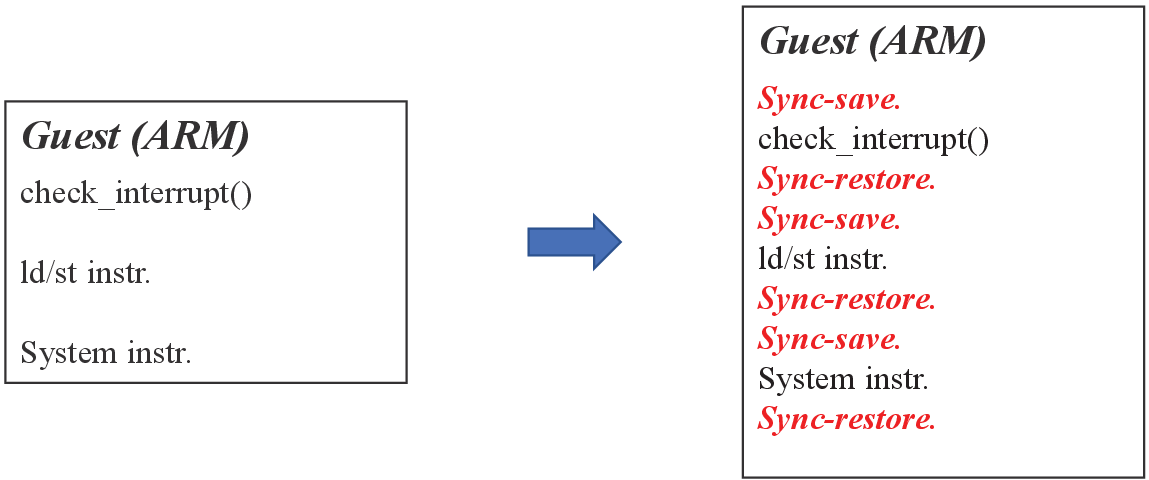}
    \caption{A \emph{basic} guest CPU state coordination.}
    \label{fig:sync_design}
\end{figure}

To reduce overall overheads caused by context switching, a basic coordination scheme is shown in Figure~\ref{fig:sync_design}.
In the rule application phase, we first perform a scan on the guest TB to check each guest instruction.
It marks the instructions in the TB, such as system-level instructions and \texttt{ld/st} instructions, that require guest CPU state coordination.
We also identify what guest CPU states these instructions may read and/or write.
Based on the information collected, the translation rule will insert codes to coordinate those CPU states when translating the corresponding instruction (see Figure~\ref{fig:sync_design}).
%Firstly, we insert coordination before and after the system-level instructions.
%Secondly, we invoke the address translation process to support the memory address translation and insert coordination before and after the memory access instructions.
%Finally, we insert coordination before and after the interrupt checking function.
%Using the design above, we overcome the coordination challenges and apply the learning-based approach to system-level DBT.

\begin{figure}[!ht]
    \centering
    \includegraphics[width=0.477\textwidth]{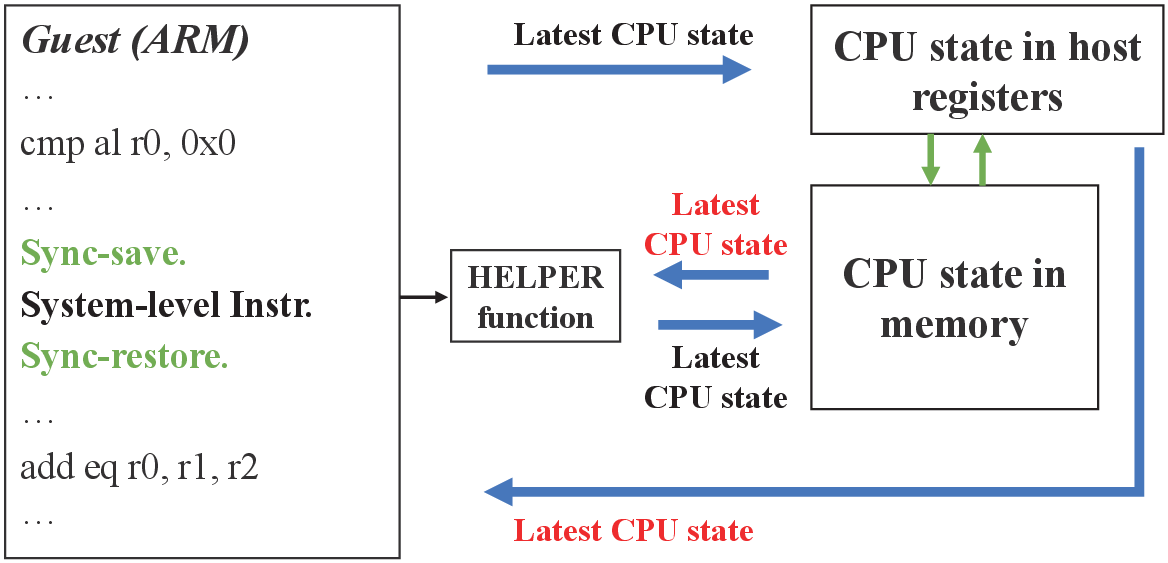}
    \caption{An example of the guest CPU state coordination for a system-level instruction.}
    \label{fig:sys_level_design}
\end{figure}

We show the example of a system-level instruction such as a \texttt{vmrs}/\texttt{vmsr} in Figure~\ref{fig:sys_level_design}.
The translation rule will insert a \emph{Sync-save} and a \emph{Sync-restore} before and after the helper function used to emulate the guest system-level instruction.
%(they are used to transfer contents between an general-propose register and a system register like \emph{FPSCR} register).
In the \emph{Sync-save} before a \texttt{vmrs} instruction, it will upload the latest guest CPU state updated by the previous rule-translated \emph{cmp al} instruction to QEMU, which allows the helper function to get the latest guest CPU state.
Similarly, if the system-level instruction is a \texttt{vmsr} instruction, a \emph{Sync-restore} will pass the latest guest CPU state from QEMU to the host registers after emulating this instruction.
%where learning-based approach maintaines them.
It allows the following instruction "\texttt{add eq}" to have the latest guest CPU state in the host registers when we apply the learned translation rule on it.%\texttt{add eq}.

However, such a naive implementation will require very frequent guest CPU state coordination and incur a lot of runtime overhead.
Thus, we propose three optimizations in the following subsections to reduce its overhead.

%------------------------------------------------------------------------------------------------------------
%------------------------------------------------------------------------------------------------------------
%\subsection{Lazy Parsing Optimization}
\subsection{Coordination Overhead Reduction}
\label{sec:sync_overhead_reduction}
% In our base design, we coordinate the CPU state between learning-based approach and QEMU for two reasons.
% On the one hand, we do this to prevent the CPU state in the host registers from being corrupted by the native codes of QEMU.
% On the other hand, we need to ensure that learning-based approach and QEMU can respectively get the latest CPU state when the translation context switches between them.
In the three scenarios mentioned in Section~\ref{sec:background_issues}, only system-level instructions will update and modify guest CPU states. 
%QEMU will read or write the guest CPU state, such as emulating system-level instructions.
For address translation and interrupts, 
%QEMU does not actually read or write the guest CPU state, and the effect of our 
guest CPU state coordination is required only to prevent these states from being corrupted during a context switch.
However, even if we only maintain those CPU states that will be modified in a context switch, our experimental results show that it still requires \instrPerSync~instructions in each context switch, i.e., it only yields very modest overhead reduction.

In addition, in the rule-based binary translation\footnote{Since the learning-based DBT uses translation rules for the binary translation, we also call it \emph{rule-based} binary translation.}, it may maintain several components of the CPU state in one register, but are maintained in separate memory locations by QEMU.
A typical example is the bit-wise condition codes and flags, which are maintained in \emph{one} \texttt{condition-code register} (CCR) in the rule-based translation since both x86 and ARM have such registers. 
But each condition code bit is kept in a different memory location in QEMU.
We call this type of CPU state "\emph{one-to-many} CPU state".
In a Sync-save operation, it will need to parse the host CCR register and store each condition code in a different memory location using several store operations.
However, if QEMU does not use them in its emulation, we can save the  CCR register in one memory location using only one store operation and restore them afterward with only one load operation.
In this way, we can reduce the total number of memory operations in Sync-save and Sync-restore operations.

%Based on the above observations, we propose a lazy parsing optimization.
%Based on the above observation, we propose a coordination overhead reduction scheme.
Based on the above observation,
depending on the types of the instructions collected in the lightweight parsing of a translation block, if a Sync-save operation involves an one-to-many state, we store the content of the host register that maintains the state to one memory location.
In the case of the CCR register, unless QEMU needs to access any of the condition codes for emulation, we need not store them in separate memory locations designated by QEMU.
%We delay the parsing the CPU states and saving them to QEMU until QEMU actually needs them.
%In the scenarios QEMU does not need them, we can omit the parsing phase.
As most of the Sync-save and Sync-restore operations are just to keep guest CPU state from being corrupted during context switches (see Table~\ref{tab:issue_percentage}), a significant number of memory operations can be eliminated this way.
%As long as they are rarely used by QEMU, this optimization can greatly reduce the number of actually executed instructions of the coordination.

\begin{figure}[!ht]
    \centering
    \includegraphics[width=0.477\textwidth]{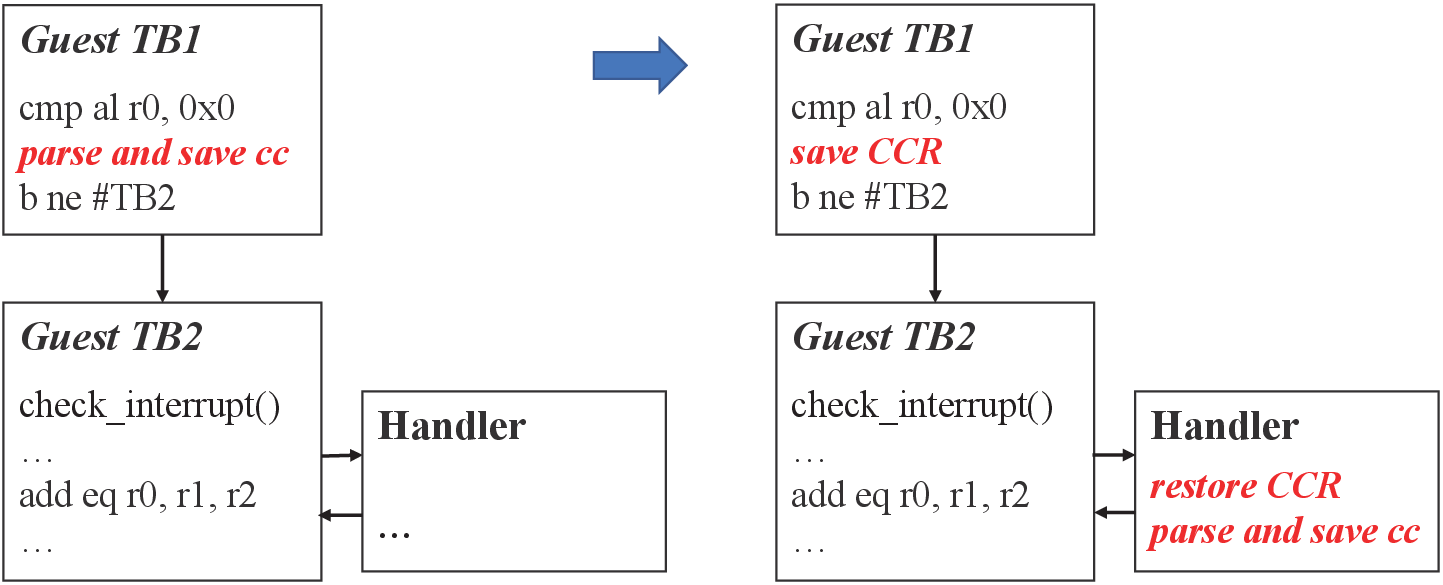}
    %\caption{Lazy parsing optimization.}
    \caption{Coordination overhead reduction.}
    \label{fig:lazy_parsing_opt}
\end{figure}

%Figure~\ref{fig:lazy_parsing_opt} illustrates the lazy parsing optimization with a typical one-to-many CPU state, i.e., condition code.
Figure~\ref{fig:lazy_parsing_opt} shows an example that involves condition codes.
It includes a 
\texttt{check\_interrupt()} in TB2 that requires access to the condition codes.
However, interrupts only occur very rarely.
%Interrupts sometimes occurs between TB1 and TB2,
%and the interrupt handler is executed at the very beginning of TB2, which requries to read the latest condition codes.
In this case, we save the content of CCR to a memory location at the end of TB1.
%Using lazy parsing, it can be done directly by saving the value of the CCR to memory at the end of TB1.
If an interrupt has been triggered at the beginning of TB2, we restore the content of CCR, parse CCR and save the condition codes separately to their designated memory locations in QEMU.
This is needed only when an interrupt occurs and the condition codes are actually needed. 
From our experimental results, we find such interrupts occur very rarely in most applications.
For example, it occurs only 0.0001\% per guest instruction in \spec.
%(Due to space limitations, we do not give the details.).
A significant number of memory operations can thus be avoided.
%Thus, we only need to save the value of CCR to memory without parsing them in most cases.

\begin{figure}[!ht]
    \centering
    \includegraphics[width=0.477\textwidth]{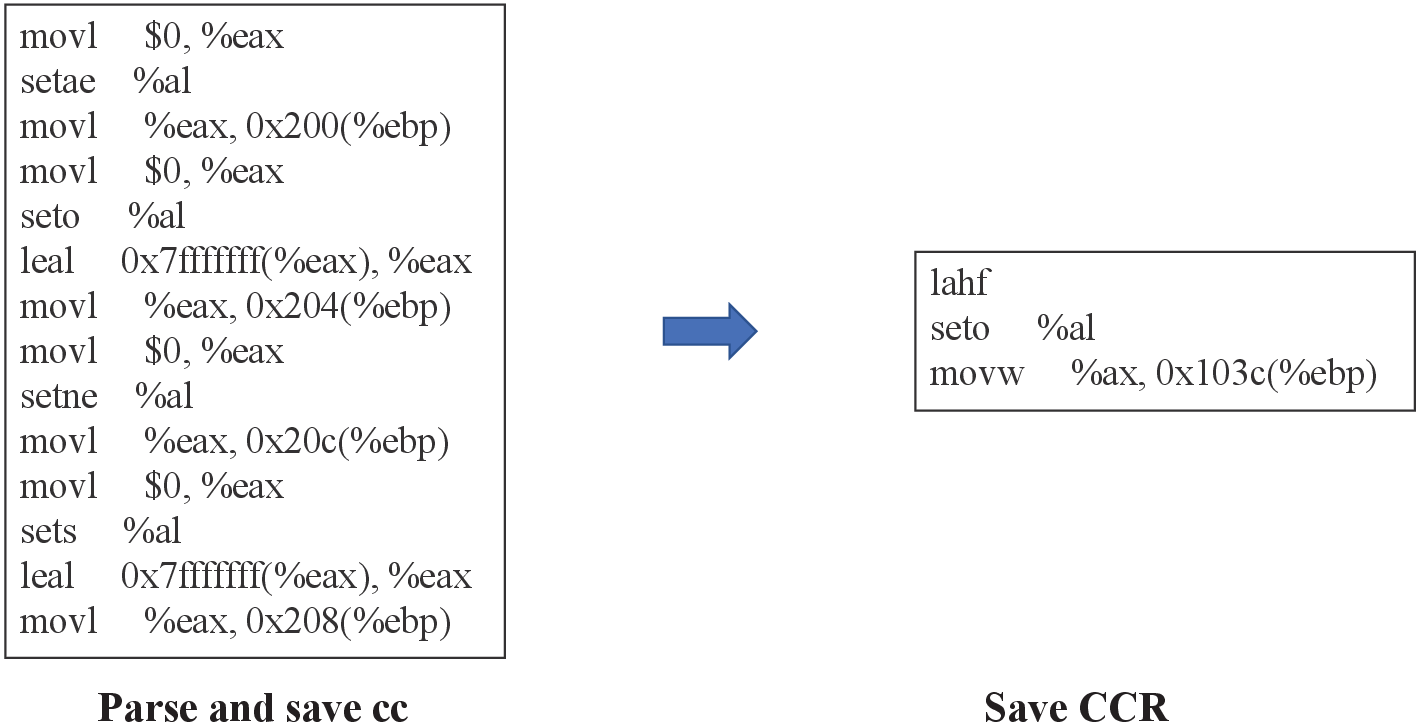}
    %\caption{Effect of lazy parsing optimization.}
    \caption{Effect of coordination overhead reduction.}
    \label{fig:lazy_parsing_effect}
\end{figure}

% The example above demonstrates the application of lazy parsing optimization to the coordination design for interrupts.
Similar optimization can be applied to Sync-save and Sync-restore operations for system-level instructions and address translation as well.
%It significantly reduces the instruction number and thus the overhead of one-to-many CPU state coordination.
As the example shown in Figure~\ref{fig:lazy_parsing_effect}, it needs about 14 instructions to parse the Eflags register (the CCR of x86) and save the condition codes to QEMU.
After applying the optimization, only 3 instructions are actually needed, with a saving of (14-3)/14 = 78\%.

%\begin{figure}[!ht]
%    \centering
%    \includegraphics[width=0.477\textwidth]{figures/lazy_parsing_effect}
    %\caption{Effect of lazy parsing optimization.}
%    \caption{Effect of coordination overhead reduction.}
%    \label{fig:lazy_parsing_effect}
%\end{figure}

%------------------------------------------------------------------------------------------------------------
%------------------------------------------------------------------------------------------------------------
\subsection{Coordination Elimination}
\label{sec:sync_elimination}
%Although the overhead of each coordination is reduced by the lazy parsing optimization, we still need to do CPU state coordination frequently.
Even after the overhead of each guest CPU state coordination is reduced by the optimization mentioned above, there are still cases where such coordination operations can be eliminated all together.
%and it achieves the same effect at the same time.
We propose three such optimizations in this section.

%------------------------------------------------------------------------------------------------------------
%------------------------------------------------------------------------------------------------------------
\subsubsection{Redundant Sync-restores Elimination}

% \textcolor{red}{In the basic rule-based coordination scheme described in Section~\ref{sec:basic_design}, if the  guest CPU state is maintained in QEMU and if a guest instruction needs to access such information, it inserts Sync-restore to pass such information to the host registers for the translated binaries in the code cache.
% If a host  instruction will update the CPU state such as condition codes (e.g. \texttt{cmp} or \texttt{add}) which is inconsistent with the semantics of its corresponding host instruction, it is allowed in a constrained rule [??ATC19]. However, an followed instruction that will use the CPU state (e.g. \texttt{b ne}), 
% 	%the basic rule-based design will insert a Sync-save after the first instruction and a Sync-restore before the second instruction.
% 	the basic rule-based design will insert a Sync-restore before the followed instruction to get the correct CPU state.
% 	However, in this case, as the restored CPU state will be used immediately in the second instruction, the Sync-save for the Sync-restore for the second instruction are no longer necessary.
% 	%after performing the first sync-restore, the latest CPU states are already in the host registers, and the following sync-restore is no longer necessary.
% }

There are conditionally-executed instructions in some ISAs, which are executed based on whether the condition specified by the instruction is satisfied or not.
For example, \texttt{add eq} is a conditional instruction in ARM-v7 that depends on the condition code \texttt{Z}.
Only when the condition code \texttt{Z} is set, this instruction will be executed.
In a DBT system, when a \emph{guest} conditional instruction is translated, a \emph{host} comparison instruction (e.g., \texttt{cmpl} in x86) is used to determine whether the current condition is satisfied or not.
This will cause the guest CPU state maintained in the host registers to be corrupted.
In the rule-based approach, as a host comparison instruction will change its CPU state that may be inconsistent with the semantics of its corresponding guest instruction, it is allowed only in a \emph{constrained rule}~\cite{DBLP:conf/usenix/SongWYZZ19}.
The basic rule-based design will insert a Sync-restore after the comparison instruction to maintain the correct guest CPU state.
However, if we encounter consecutive conditional instructions that depend on the same condition, we only need to restore the guest CPU state once at the first conditional instruction, and the remaining instructions can be translated normally without the additional comparison instructions and Sync-restore operations. 
%since the condition has been judged at the first instruction.
%\textcolor{red}{In the basic rule-based coordination scheme described in Section~\ref{sec:basic_design}, if the  guest CPU state is maintained in QEMU and if a guest instruction needs to access such information, it inserts Sync-restore to pass such information to the host registers for the translated binaries in the code cache.
%If a guest instruction will update the CPU state such as condition codes (e.g. \texttt{cmp} or \texttt{add}), and is followed by an instruction that will use the CPU state (e.g. \texttt{b ne}), 
%%the basic rule-based design will insert a Sync-save after the first instruction and a Sync-restore before the second instruction.
%the basic rule-based design will insert a Sync-save and a Sync-restore before and after each instruction.
%However, in this case, as the updated CPU state will be used immediately in the second instruction, the Sync-save for the first instruction and the Sync-restore for the second instruction are no longer necessary.
%%after performing the first sync-restore, the latest CPU states are already in the host registers, and the following sync-restore is no longer necessary.
%}
%An obvious optimization is to remove such redundant coordinations among the instructions in a TB. %sync-restore optimization to omit the unnecessary sync-restore in this scenario.
%In this optimization, 
To do this, in each TB, it first checks if there are such instructions that update the CPU state and the conditional instructions that depend on the updated CPU state in the TB.
%read-CPU-state instruction sequence after a define-CPU-state instruction.
Next, it keeps the first Sync-restore before the first instruction in the TB that uses the CPU state, and eliminates the other Sync-restores until it reaches an instruction that requires a Sync-save operation or reaches the end of the TB.

\begin{figure}[!ht]
    \centering
    \includegraphics[width=0.477\textwidth]{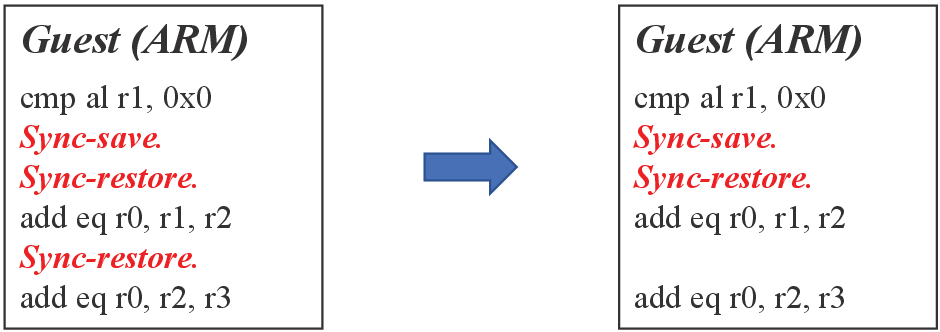}
    \caption{Coordination-restore optimization.}
    \label{fig:sync_restore_opt}
\end{figure}

As the example shown in Figure~\ref{fig:sync_restore_opt},
we assume that after the instruction "\texttt{cmp al}", some situation (e.g., a system-level instruction) requires the Sync-save operation to save the CPU state to QEMU.
At the same time, the next few instructions "\texttt{add eq}" need to use the CPU state.
The rule-based translation will have a Sync-restore for each of such instructions. 
%In the base design, before each instruction \emph{add eq}, it needs to restore the latest CPU state from QEMU to host registers.
But in this case, only the first Sync-restore is needed, and the rest of the Sync-restores and the translated comparison instructions can be eliminated.
%After optimization, we check this situation and only do sync-restore once.
Through this optimization, a significant amount of coordination overhead due to the rule-based translation can be eliminated.

%------------------------------------------------------------------------------------------------------------
%------------------------------------------------------------------------------------------------------------
\subsubsection{Optimization for Consecutive Memory Operations}

In the case of consecutive memory-access instructions, we can use a similar coordination elimination scheme to reduce redundant coordination.
Because of the need to emulate address translation in QEMU  at the system level, a Sync-save operation is inserted before and a Sync-restore after each memory access instruction during the rule-based translation.
Apparently, if there are consecutive memory-access instructions in a TB, the intermediate coordination among those instructions can be removed.
%We propose a continuous ld/st optimization to remove the redundant coordinations in this scenario.
In this case, it first checks if it has a sequence of consecutive memory-access instructions in the TB.
If it has, it keeps the Sync-save operation before the first memory-access instruction and the Sync-restore operation after the last memory access instruction in the sequence.
The rest of the intermediate Sync-save and Sync-restore operations in the sequence can be removed.

\begin{figure}[!ht]
    \centering
    \includegraphics[width=0.477\textwidth]{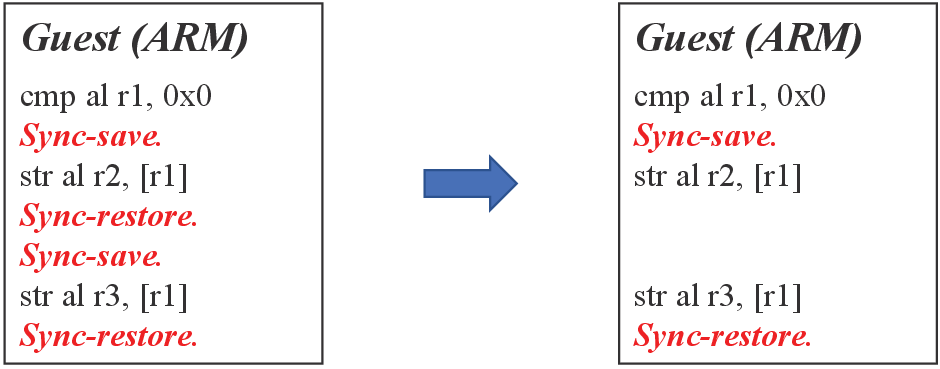}
    \caption{Optimization on consecutive ld/st instructions.}
    \label{fig:ldst_opt}
\end{figure}

As the example shown in Figure~\ref{fig:ldst_opt},
there are two consecutive memory-access instructions \texttt{str} after the instruction "\texttt{cmp al}". 
The rule-based translation will insert two pairs of coordination in this situation.
%After continuous ld/st optimization, we check continuous memory access instruction sequences and only do sync-save once at the beginning and sync-restore once at the end of these sequences.
After applying the redundant coordination elimination, only a pair of Sync-save and Sync-restore are needed.

%------------------------------------------------------------------------------------------------------------
%------------------------------------------------------------------------------------------------------------
\subsubsection{Inter-TB Optimization}

As the unit of translation in a DBT is a guest TB, it needs to save the CPU state at the end of each TB and switch back to QEMU because the following TB may yet to be translated, or it may need to do a \texttt{check\_interrupt} before the next TB.
However, \emph{block chaining} is a common optimization~\cite{DBLP:conf/aPcsac/WangHKNBYW07}\cite{DBLP:conf/cgo/HongHYWHLWC12} that chains multiple TBs in the code cache together without context switching back to QEMU after each TB.

In this case, if the \emph{first} guest instruction of the next TB will update a CPU state without using the CPU state defined in the previous TB, 
%the need of before using it for the first time, i.e., the next TB defines a new CPU state and does not use the corresponding CPU state defined by the previous TB, 
we will need neither a Sync-save operation at the end of the previous TB nor a Sync-restore at the beginning of the current TB.
%Unfortunately, interrupts may occur at any time, and the dynamic binary translator needs to execute the TBs translated from the interrupt handler, thus changing the expected execution flow of the program.
%Therefore, it cannot be sure which TB will be the next.
After an in-depth analysis of the execution flow in application codes, we find that if a series of TBs are chained together such an execution flow can be analyzed.
In other words, in a code cache that uses block chaining similar to one used in QEMU, there is an opportunity to eliminate the Sync-save at the end of the current TB and the Sync-restore operation at the beginning of the next TB if the next TB will update its CPU state before using it.

Based on the above observation, we propose an inter-block optimization.
It first checks if the current TB will jump to a TB in the code cache via block-chaining. 
Next, for each CPU state that needs to be coordinated, check the next TB to see if there is an instruction using the CPU state before it is updated by an earlier instruction in the TB.
%If not, we can be sure that instructions in the next TB do not use this CPU state from the current TB and 
If not, we can omit the Sync-save operation at the end of the current TB and the Sync-restore operation at the beginning of the next TB.

\begin{figure}[!ht]
    \centering
    \includegraphics[width=0.477\textwidth]{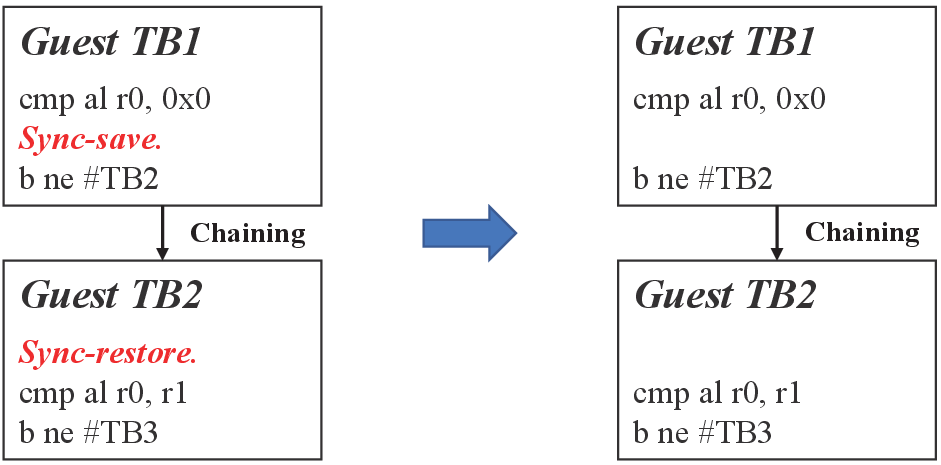}
    \caption{Inter-TB optimization.}
    \label{fig:inter_tb_opt}
\end{figure}

Figure~\ref{fig:inter_tb_opt} shows such an example.
After the rule-based translation, the instruction "\texttt{cmp al}" in TB1 will update the condition codes.
It needs to do a Sync-save operation for the condition codes to be used in the following TB.
In addition, the following TB2 needs to do a Sync-restore operation to obtain the correct condition codes.
After traversing the block chain, it is found that TB1 jumps to TB2 in the chain.
Furthermore, the instruction "\texttt{cmp al}" in TB2 updates the condition codes before the instruction "\texttt{b ne}" uses the condition codes.
We can thus eliminate the Sync-save operation in TB1 and the Sync-restore operation in TB2. 
%make sure that the next TB of TB1 is TB2, and TB2 defines new condition codes and does not need the condition codes defined by TB1.
%In this situation, we can omit the condition code coordination at the end of TB1.

% In addition to condition codes, similar checks can also be performed and applied inter-TB optimization on other CPU states, such as general-purpose registers.
% With this optimization, we can reduce redundant coordinations and gain a better code quality.

%------------------------------------------------------------------------------------------------------------
%------------------------------------------------------------------------------------------------------------
\subsection{Instruction Scheduling}
\label{sec:instruction_scheduling}
We can also use instruction scheduling to reduce more redundant coordination.
In this section, we present two scenarios, called \emph{define-before-use scheduling} and \emph{interrupt scheduling}, that can further reduce such coordination.

%------------------------------------------------------------------------------------------------------------
%------------------------------------------------------------------------------------------------------------
\subsubsection{Define-Before-Use Scheduling}

Within a TB, the instruction that uses the CPU state such as condition codes may be several instructions behind the instruction that updates such CPU state.
There may be other instructions in between also. 
Some of them may even be system-level instructions or ld/st instructions that require QEMU's intervention.
In the rule-based translation, a Sync-save operation will be inserted after the "update/define" instruction and a Sync-restore will be inserted before the "use" instruction.
But, if there is no instruction in between that is data-dependent on either instruction, we can schedule the two instructions next to each other to avoid a CPU state coordination.
%In this way, there is no need to coordinate the CPU state.
We call such an instruction scheduling scheme the \emph{define-before-use scheduling} scheme.
%By collecting the data dependence information during TB traversal process, we can reorder the instructions between defining and using CPU state if dependence analysis permits.
% Define-before-use scheduling is to reorder the instructions between defining and using CPU state if dependence analysis permits.
% We marking the information about what CPU states the instructions define and use in the TB during TB traversal process.
% Afterward, we look for the instruction pair in the TB with a define-use relationship on the same CPU state.
% Then, we analyze whether the instructions between this instruction pair depend on the instruction pair.
% If not, we move the instructions to the location just before this instruction pair.
%In this way, the coordination can be omited that would otherwise be required between the instructions.

\begin{figure}[!ht]
    \centering
    \includegraphics[width=0.477\textwidth]{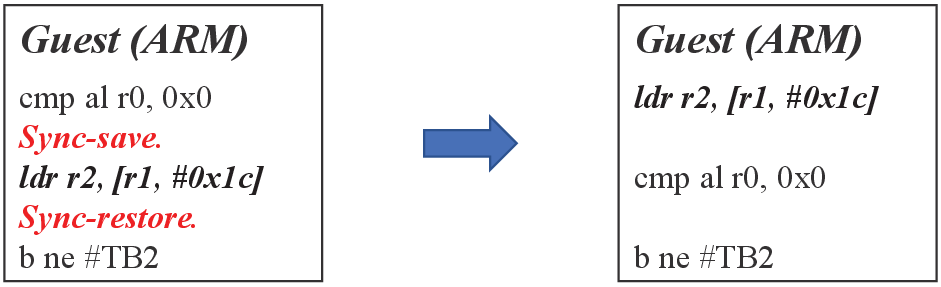}
    \caption{Define-before-use scheduling.}
    \label{fig:scheduling1}
\end{figure}

Figure~\ref{fig:scheduling1} shows such an example.
In this example, the instruction "\texttt{cmp al}" updates the condition codes, and the instruction "\texttt{b ne}" uses them.
During the rule-based translation, it will insert a Sync-save operation and a Sync-restore operation before and after the memory-access instruction \texttt{ldr}.
However, the \texttt{ldr} instruction is not dependent on the "\texttt{cmp al}" instruction nor the "\texttt{b ne}" instruction.
By scheduling both instructions together after the \texttt{ldr} instruction, there is no longer a need to coordinate these condition codes before and after the \texttt{ldr} instruction.

%------------------------------------------------------------------------------------------------------------
%------------------------------------------------------------------------------------------------------------
\subsubsection{Interrupt-driven Scheduling}

QEMU handles interrupts at the system level by inserting an interrupt-check function at the beginning of each TB.
In theory, we can place the interrupt-check function in any other location in the TB.
In the rule-based translation scheme, we insert CPU state coordination before and after each interrupt-check function.
If the TB has memory-access instructions, we also need such coordination for each such instruction to prevent possible inconsistencies.
Similar to the instruction scheduling scheme described earlier, if we can schedule the interrupt-check function close to the memory-access instructions, we can eliminate those redundant coordination.
As the memory-access instructions appear quite frequently while interrupts rarely occur, 
there are ample opportunities to cut down such redundant coordination.
We call such an approach \emph{interrupt-driven scheduling}.
It is particularly effective if block chaining is applied to TBs in the code cache.

%Interrupt scheduling is to merge the interrupt check function to the nearest memory access instruction at the same block if the last TB jumps by block-chaining.
%First, finding the first ld/st instruction of the TB.
%Then, checking whether the instructions between the interrupt check function and the ld/st instruction use the CPU state defined by the interrupt handler.
%If not, moving the interrupt check function before the ld/st instruction without disrupting the correct execution.
%In this way, only one pair of coordination is needed instead of two pairs.

\begin{figure}[!ht]
    \centering
    \includegraphics[width=0.477\textwidth]{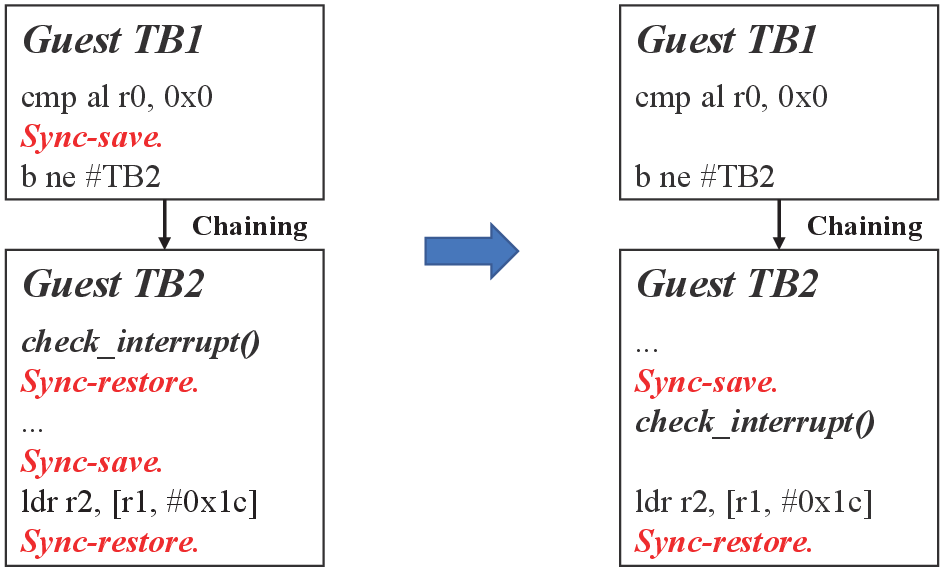}
    \caption{Interrupt scheduling.}
    \label{fig:scheduling2}
\end{figure}

Figure~\ref{fig:scheduling2} shows such an example.
%In this example, TB1 jumps to TB2 by block-chaining.
Initially, CPU state coordination operations will be inserted both at the interrupt-check function and the memory-access instruction \texttt{ldr}.
%However, the instruction "\texttt{sub al}" does not use any condition codes updated by the interrupt handler.
%\textcolor{red}{(????There is no "sub al" instruction in the example in Figure 14??? Need to revise Figure 14!!)}
In fact, we can move the interrupt-check function to the front of the memory-access instruction \texttt{ldr}.
This scheduling will not affect the interrupt handling and can reduce the coordination from two pairs to one.

%------------------------------------------------------------------------------------------------------------
%------------------------------------------------------------------------------------------------------------
\subsection{Optimization Interaction}
The optimizations described above may change the corresponding host basic blocks in different ways.
Therefore, we set different priorities among these optimizations and coordinate them for the best results.
Redundant sync-restores elimination and optimization for consecutive memory operations mentioned in~\ref{sec:sync_elimination}, whose trigger conditions are orthogonal, simply reduce intra-block coordination and are applied first.
Then, if the conditions of inter-TB optimization mentioned in~\ref{sec:sync_elimination} are met, we apply it to remove the coordination at the end of the block and the beginning of the next block.
Otherwise, we use the coordination overhead reduction mentioned in~\ref{sec:sync_overhead_reduction} at the end of the block.
Finally, we apply the instruction scheduling, changing the instruction order and further eliminating coordination.
Note that this optimization order will not activate any previous optimizations.

%-------------------------------------------------------------------------------
\section{Evaluation}
%-------------------------------------------------------------------------------
\label{sec:evaluation}

In this section, we evaluate our design and try to answer the following questions:
(1) Using the rule-based translation approach at the system level with the proposed optimization to reduce redundant guest CPU state coordination, how much performance improvement can we achieve compared to state-of-the-art systems like \qemu{}?
(2) How do various optimizations proposed in the paper affect the overall performance?
(3) How does the performance of our approach compare to that of the native execution?
%(4) What is the dynamic translation coverage of the rule-based approach when applied to the system-level emulation? 
(4) What about its performance improvement for real applications?  

% \begin{itemize}

% \item Using the rule-based translation approach at the system level with the optimization proposed to reduce redundant guest CPU state coordination, how much performance improvement can we achieve compare to the state-of-the-art systems like \qemu{}?

% \item What is the dynamic translation coverage of the rule-based approach when applied to the system-level emulation?

% \item How do various optimizations proposed in the paper affect the overall performance?

% \item How does the performance of our approach compare to that of the native execution?

% \end{itemize}

%------------------------------------------------------------------------------------------------------------
%------------------------------------------------------------------------------------------------------------
\subsection{Experimental Setup}

We have implemented a system-level rule-based DBT prototype based on \qemu{}. 
We take ARM-v7 as the guest ISA and Intel x86 as the host ISA.
The translation rules used are the same parameterized translation rules used in~\cite{DBLP:conf/micro/JiangDZSWYZ20}. The prototype is run on an Intel Xeon E5-2680 v4 machine with 2 cores, 56 threads and 126GB DRAM.
The host OS is a 32-bit Ubuntu 14.04 with Linux 3.13.
The guest OS is a 32-bit unmodified Linux system with kernel 4.4.0.
We compile the \spec~using GCC-4.8 with -O2 optimization level and statically linking, and run the \emph{ref} input of \spec~\footnote{The rule-based approach also supports the translation of floating-point instructions. Due to the space constraint, the floating-point applications in SPEC 2006 are not listed here. When these applications are included, our approach can achieve an average of 1.92X speedup over \qemu, instead of 1.36X speedup for only \spec.} on the guest OS.
%\blue{
To better understand the performance of the rule-based approach, we also evaluate several real-world applications that include \emph{memcached, sqlite, fileIO, untar} and \emph{cpu-prime}, which are widely used in other system research~\cite{DBLP:conf/vee/HuangZLNWL21, DBLP:journals/corr/abs-2201-09652}.
%The descriptions are listed in Table~\ref{tab:app}.
%}
We run each benchmark ten times and take the average to reduce the effect of fluctuation.
To measure the performance speedup, we use the execution time on \emph{unmodified} \qemu{} as the baseline.

%\begin{table}[!ht]
%	\centering
%	\huge
%	\caption{\blue{The Description of Real-World Applications.}}
%	\renewcommand\arraystretch{3}
%	\label{tab:app}
%	\scalebox{0.45}{
%	\begin{tabular}{|c|c|}
%		\hline
%		\textbf{Name} & \textbf{Description} \\ \hline
%		Memcached   & \parbox[c]{16cm}{Memcached v1.6.19 running the memslap with 16 threads and sending 100000 requests.}		\\ \hline
%		Sqlite     & \parbox[c]{16cm}{Sqlite is used to create 10000000 insertions, each inserting 3 integer and 1 fp values.} 	\\ \hline
%		FileIO     & \parbox[c]{16cm}{FileIO test in sysbench v0.5. Randomly reading 6000 times and write 4000 times with 128 files of 4MB each. Block size is 16KB.} 	\\ \hline
%		Untar     & \parbox[c]{16cm}{Untar extracting a 531MB tarball using tar utility.} 	\\ \hline
%		CPU-prime     & \parbox[c]{16cm}{CPU test in sysbench v0.5. Calculating prime numbers up to the max prime 10000.} 	\\ \hline
%	\end{tabular}}
%\end{table}

%------------------------------------------------------------------------------------------------------------
%------------------------------------------------------------------------------------------------------------
\subsection{Overall Performance}

To study the effectiveness of our approach, we collect the performance data on \qemu{} that includes the un-modified \qemu{}, the rule-based implementation of the system level on \qemu{}, and its optimized version with the three optimizations described in Section \ref{sec:design_opt}~(marked as "Full Opt." in the following figures). 
%For a better comparison of the performance, we also collect the performance data of the rule-based enhanced  \qemu{} in the user mode~\cite{DBLP:conf/micro/JiangDZSWYZ20}\cite{DBLP:conf/usenix/SongWYZZ19}\cite{DBLP:conf/asplos/WangMZY18}.

\begin{figure}[!ht]
	\centering
	\includegraphics[width=0.477\textwidth]{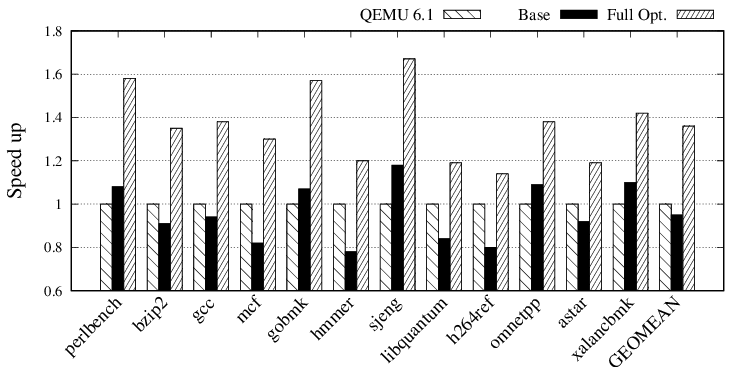}
	\caption{Performance of \spec~running in system mode on un-modified \qemu{}, the un-optimized rule-based implementation of \qemu{}, and the optimized version of the rule-based implementation.}
	\label{fig:speedup}
\end{figure}

%The performance data are shown in Figure~\ref{fig:speedup}.
As the results in Figure~\ref{fig:speedup} show, the un-optimized
rule-based implementation of \qemu{} has a \baseSlowdownQemu{} slowdown compared to \qemu{}, i.e., it is actually slower than QEMU 6.1 running in system mode.
The main reason is the CPU state coordination overhead explained in Section~\ref{sec:design_opt}.
However, after the three optimizations are applied to reduce coordination overheads, the optimized rule-based \qemu{} running at the system level can achieve a \optSpeedup X speedup.
% Also, compared to the performance data of the benchmarks running in the user mode, the performance in the system mode has a similar trend as shown in Figure~\ref{fig:speedup}.

%The reason for the poor performance of our base design is that the coordination proportion is high and the overhead of each coordination is large.
Based on the data in Table~\ref{tab:issue_percentage}, an average of \syncOpt~guest instructions (which include system-level instructions, memory-access instructions and interrupt checks) will require CPU-state coordination operations.
Moreover, each coordination operation will introduce around \instrPerSync~host instructions.
%For the base design, the performance of learning-based method 
From Table~\ref{tab:issue_percentage}, for some benchmarks such as \emph{mcf} and \emph{h264ref}, the percentages of instructions that require such coordinations run as high as 62.17\% and 64.49\%, respectively.
Their performance also suffers the most compared to other benchmarks as shown in Figure~\ref{fig:speedup}.
%is slower than that of \qemu{}.
%The major reason is that the coordination proportion of these benchmarks is high.
%For example, the coordination proportion of \emph{mcf} is 62.17\% and that of \emph{h264ref} is 64.4\% as the data in Table~\ref{tab:issue_percentage} shown.

\begin{figure}[!ht]
	\centering
	\includegraphics[width=0.477\textwidth]{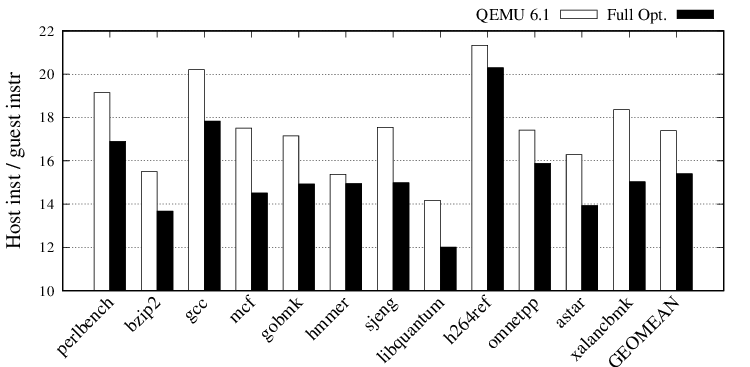}
	\caption{Average number of host instructions needed to translate a guest instruction in un-modified \qemu{}, and in the optimized rule-based implementation.}
	\label{fig:trans_quality}
\end{figure}

After the three optimizations are applied to remove redundant coordination operations, the percentage of instructions that require coordination is reduced to \schedulingSyncOpt{}, and the number of host instructions required in each coordination operation also goes down to only around 3 host instructions (as shown in Figure~\ref{fig:lazy_parsing_effect}).
To further understand the effect of the rule-based approach, we also collect the average number of host instructions needed to translate a guest instruction.
The data are shown in Figure~\ref{fig:trans_quality}.
As the data show, \qemu{} in system mode requires an average of around \baseTransNum{} host instructions for each guest instruction, while the optimized rule-based implementation requires an average of \optTransNum{} host instructions - a reduction of around 11.44\%.
%In Section~\ref{sec:evaluation_opt}, we will describe the effect of sequentially applying the optimizations to the base design in detail.

%The slowdown factor of a system-level emulation compared to the native execution of a program is an important factor in designing an emulator.
%We collect such performance data for both un-enhanced \qemu{} and our rule-based optimized version of \qemu{}. Due to the space constraint, the detailed data are not given out. 
%Our rule-based approach achieves an average slowdown of \optSlowdown X while \qemu{} has an average of \qemuSlowdown X slowdown.

%We observe that in the base design, the speedup of \emph{sjeng} and \emph{perlbench} reaches respectively 1.21X and 1.18X, and in the optimized design, the speedup even reaches 1.67X and 1.58X.
%But for \emph{gcc}, performance improvement brought by the base design is low, only 1.03X.
%For some benchmarks such as \emph{mcf} and \emph{h264ref} the performance of the base design even drops, only 0.95X and 0.84X that of QEMU 6.1, which means in these benchmarks, the advantages brought by the learning-based approach can no longer make up for the overhead.
%This is because there are more CPU state that need to be coordinated at the end of TB in these benchmarks.
%As shown in Table~\ref{tab:issue_percentage}, \emph{mcf} and \emph{h264ref} respectively have 62.17\% and 64.4\% instructions introduce coordination operations.
%However, if we use the three optimizations, the percentage of \emph{mcf} and \emph{h264ref} drops to  26.95\% and 25.90\% and the speedup reaches 1.30X and 1.14X.
%The observation above demonstrates that there are indeed many inefficiencies in our base design.

To identify the performance bottleneck, we count the number of instructions in the host basic blocks and group the instructions by their functionality.
Based on our analysis, one of the major bottlenecks is in the address translation.
Since QEMU needs to emulate MMU behavior for each memory access in system mode, it involves about 20 host instructions for each translated memory instruction on average.
This shows that the address translation incurs very high overheads, and it will be the focus for further optimization in our future work.

%------------------------------------------------------------------------------------------------------------
%------------------------------------------------------------------------------------------------------------
\subsection{Impact of Coordination Optimizations}
\label{sec:evaluation_opt}

\begin{figure}[!ht]
    \centering
    \includegraphics[width=0.477\textwidth]{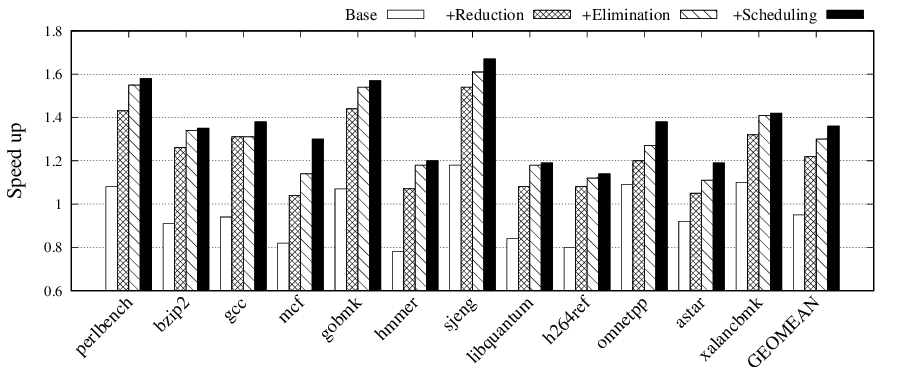}
    \caption{Cumulative performance improvement after adding each proposed optimization.}
    \label{fig:opt_speedup}
\end{figure}

To understand the performance impact of each coordination optimization, we evaluate the \emph{cumulative} performance improvement after adding each optimization.
%of three optimizations separately.
The results are shown in Figure~\ref{fig:opt_speedup}.

In the figure, `\textit{Base}' marks the performance of the un-optimized version as described in Section~\ref{sec:basic_design}.
`\textit{+ Reduction}' marks the performance after adding the optimization to reduce the number of host instructions in a coordination operation as described in Section~\ref{sec:sync_overhead_reduction}.
%that only the reduction optimization is applied based on the based design.
`\textit{+ Elimination}' marks the performance after further adding the optimization that eliminates redundant coordinations as described in Section~\ref{sec:sync_elimination}.
%optimization is applied after the reduction one is used.
`\textit{+ Scheduling}' marks the performance after further applying instruction scheduling as described in Section~\ref{sec:instruction_scheduling}.
The baseline is the performance of the unmodified \qemu{}.

As the data show, it achieves 1.22X speedup after the reduction optimization is applied.
After adding the optimization that eliminates redundant coordination operations, the cumulative performance improvement is 1.30X.
When all optimizations are applied, we achieve 1.36X overall speedup.  

\begin{figure}[!ht]
    \centering
    \includegraphics[width=0.477\textwidth]{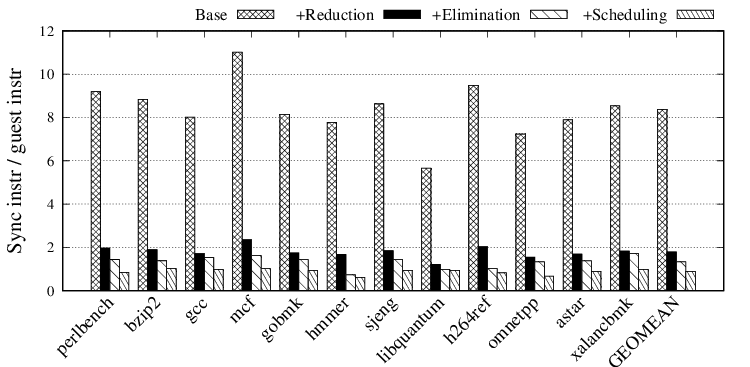}
    %\caption{Average number of instructions for coordination respectively brought by translating a guest instruction after lazy parsing, elimination and scheduling added.}
    \caption{Average number of host instructions per guest instruction for coordination after three optimizations are applied.}
    \label{fig:sync_instr}
\end{figure}

To further understand the performance impact of these three optimizations, we also calculate the average number of host instructions needed for coordination per guest instruction.
It is calculated by the following formula:$$ sync\_instr~per~guest\_ins = \frac{sync\_num*sync\_overhead}{guest\_num}$$
In the formula, \emph{sync\_num} is the total number of coordination operations, \emph{sync\_overhead} is the average number of host instructions used in a coordination operation, and \emph{guest\_num} is the total number of translated guest instructions. 
The results are shown in Figure~\ref{fig:sync_instr}.

Because we can reduce the number of host instructions in a coordination operation from 14 instructions to 3 instructions as shown in Figure~\ref{fig:lazy_parsing_effect},
%Since the reduction optimization reduces the coordination overhead from 14 to 3, 
the average number of host instructions for coordination per guest instruction is reduced from \baseSync{} to \lazySync{}.
After the elimination of redundant coordination operations, the number of host instructions for coordination per guest instruction is further reduced to \eliminationSync.
When the instruction scheduling is finally applied, that number is eventually dropped to \schedulingSync{}.

%------------------------------------------------------------------------------------------------------------
%------------------------------------------------------------------------------------------------------------
\subsection{Comparison to Native Execution}

\begin{figure}[!ht]
	\centering
	\includegraphics[width=0.477\textwidth]{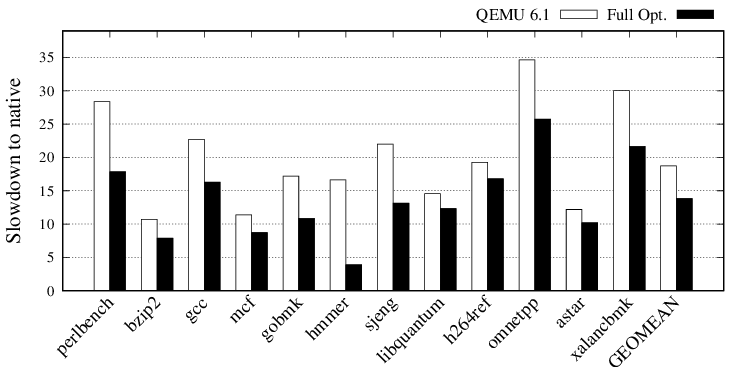}
	\caption{Slowdown factors of the system-level emulation on un-modified \qemu{} and the fully-optimized rule-based \qemu{} compared to the native execution using \spec~(lower is better).}
	\label{fig:slowdown}
\end{figure}

The slowdown factor of a system-level emulation compared to the native execution of a program is an important factor in designing an emulator.
%^The performance of native execution is an important factor in a system-level emulator.
%We thus collect the performance data of our design for native execution.
We collect such performance data for both un-modified \qemu{} and our rule-based optimized version of \qemu{}.
The data are shown in Figure~\ref{fig:slowdown}.
Compared to \qemu{}, our rule-based optimized version of \qemu{} achieves an average slowdown of \optSlowdown X while \qemu{} has an average of \qemuSlowdown X slowdown.

\subsection{Performance on Real-World Applications}

\begin{figure}[!ht]
	\centering
	\includegraphics[width=0.477\textwidth]{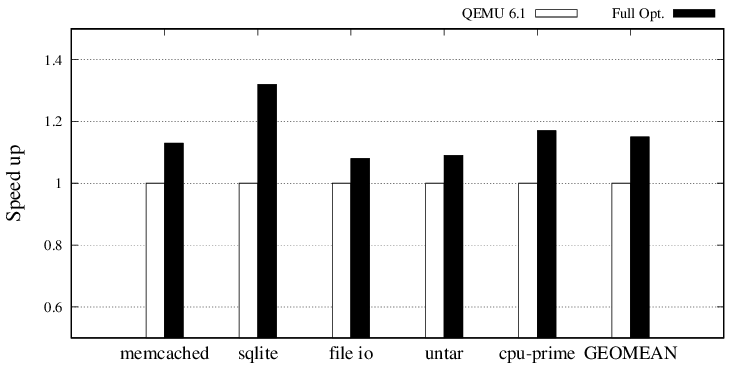}
	\caption{Speedup of real-world applications on optimized rule-based \qemu{} compared to un-modified \qemu{}.}
	\label{fig:app_speedup}
\end{figure}

%\blue{
To better understand the performance of optimized rule-based approach at the system level, we evaluate the performance with several real-world applications.
The real-world applications consists of \emph{Memcached, Sqlite, FileIO, Untar} and \emph{CPU-prime}.
As the results in Figure~\ref{fig:app_speedup} show, our design can achieve an average of \appSpeedup X speedup over \qemu{}. In these applications, \emph{FileIO} and \emph{Untar} are IO-bound applications and \emph{Memcached} is a network application. Since a lot of execution time is spent on IO or network, we can only achieve a speedup of 1.08X, 1.09X and 1.13X, respectively.
%}

%-------------------------------------------------------------------------------
\section{Related Work}
%-------------------------------------------------------------------------------
\label{sec:related}

DBT systems have attracted extensive research. Prior work includes instruction translation and optimization~\cite{DBLP:journals/esl/SalgadoGPCT17, DBLP:conf/fskd/LiuZPYBC13, DBLP:conf/csie/ChuZGL09, DBLP:conf/micro/JiangDZSWYZ20, DBLP:conf/usenix/SongWYZZ19, DBLP:conf/asplos/WangMZY18, DBLP:conf/cc/Wang21}, system-level translation~\cite{DBLP:conf/usenix/Bellard05, DBLP:conf/ppopp/WangLCWCZZ11, DBLP:conf/vee/DAntrasGGGL17, DBLP:conf/usenix/SpinkWF19}, memory-access instruction optimization~\cite{DBLP:conf/vee/ChangWHLY14, DBLP:conf/cgo/ReddiCCS07, DBLP:conf/usenix/WangYZM16}, indirect-branch optimizations~\cite{DBLP:journals/taco/DAntrasGGL16, DBLP:conf/cf/JiaYHC14, DBLP:conf/vee/JiaYWTW13, DBLP:conf/hpcc/ZhangGGHLM13}, translation of architecture-specific instructions such as SIMD instructions~\cite{DBLP:conf/date/FuWH15, DBLP:conf/IEEEpact/LiuHWFH17, DBLP:journals/spe/FuHLWH18, DBLP:journals/jsa/FuHLWH19}, translation of atomic instructions~\cite{DBLP:journals/tcad/KristienSCSSFBT20, DBLP:conf/cgo/ZhaoJCGWY21}, and more. 
In this paper, we mainly focus on system-level DBT systems.
% \cite{DBLP:conf/icpads/DingCHC11} provides a scalable multi-core parallel system-level dynamic binary translator prototype.
% \cite{DBLP:conf/icpp/ZhaoJLGWY20} goes beyond a single-node multi-core processor and extends QEMU to a distributed system using a cluster of multi-core processors.

% There have been some designs and optimizations in system-level emulation~\cite{DBLP:conf/cgo/HawkinsDBZ15}.

%There have been various system-level DBT optimizations and systems.
For system-level optimizations,~\cite{DBLP:conf/vee/ChangWHLY14} speeds up the memory address translation using embedded shadow page tables to do a direct mapping between a guest virtual address to its host physical address. 
% A novel persistent code caching framework~\cite{DBLP:conf/usenix/WangYZM16} is proposed to reduce the re-transaltion overhead for short running time applications.
% HERMES~\cite{DBLP:conf/cgo/ZhangGCCH15} uses Host-specific Data Dependence Graph (HDDG) to analyze and reduce the redundant host generated instructions.
\cite{DBLP:journals/pcs/SokolovE12} proposes a parallel system-level DBT emulator using a separate thread to optimize the translated code.
Qlet~\cite{DBLP:conf/vee/CotaC19} is a cross-ISA system-level instrumentation tool and several techniques is used to improve its performance.
%QEMU~\cite{DBLP:conf/usenix/Bellard05} is a widely-known emulator that supports both user-level and system-level DBT.
COREMU~\cite{DBLP:conf/ppopp/WangLCWCZZ11} uses one QEMU instance to emulate a multi-core system with a lightweight library for communication.
By leveraging multi-core platforms and the optimizations in LLVM, HQEMU~\cite{DBLP:conf/cgo/HongHYWHLWC12} proposed a parallel DBT system.
The kernel-level binary translation mechanism in~\cite{DBLP:conf/sosp/KediaB13} achieves a near-native performance.
\cite{DBLP:conf/pldi/DAntrasGGL17, DBLP:conf/vee/DAntrasGGGL17} translated binaries between ARM and x86 by utilizing host hardware features.
The work in~\cite{DBLP:conf/date/RokickiRD17, DBLP:journals/taco/SpinkWF16, DBLP:conf/usenix/SpinkWF19} takes advantage of hardware features to support system-level binary translation.
Captive~\cite{DBLP:conf/usenix/SpinkWF19} is a retargetable system-level DBT hypervisor.
It combines both offline and online optimizations running in a virtual bare-metal environment to deliver performance improvement.

Our work extends the rule-based DBT and applies it to the system-level DBT.
%Captive differs from our learning-based system-level emulator in several significant ways.
%Captive support different guest ISA via manually writing high-level architecture descriptions, while our approach supports different guest and host ISA via automatically learning translation rules from guest and host binaries.
%Moreover, Captive gains an exceeding performance improvement via offline and online optimizations and the executing in a virtual bare-metal environment, while our approach achieves a fine speedup mainly via the advantages of high-quality translation rules.
%\blue{
It focuses on improving the quality of instruction translation, and can be combined with other optimizations applied to system-level DBTs to further improve performance, such as memory optimizations~\cite{DBLP:conf/vee/ChangWHLY14}, parallelism~\cite{DBLP:conf/ppopp/WangLCWCZZ11} and hardware feature-based optimizations~\cite{DBLP:conf/vee/DAntrasGGGL17, DBLP:conf/pldi/DAntrasGGL17}.
%}

%-------------------------------------------------------------------------------
\section{Conclusion}
%-------------------------------------------------------------------------------
\label{sec:conclusion}

The rule-based approach using an automatic learning process to learn translation rules has shown to be effective in a DBT such as QEMU at the \emph{user level}.
To apply this approach to the \emph{system level}, this paper presents a basic design to coordinate CPU states embedded in the guest and the host instructions when switching between the execution from the code cache and the emulation in QEMU.
We address the issues critical to such a design and propose several optimization strategies to reduce such coordination overhead.
We also implement a prototype based on \qemu~to demonstrate the feasibility of such an approach.
%collect performance and related data that helps to understand those performance data. 
The experimental results show that our design is quite efficient. 
Compared to QEMU 6.1, our fully optimized system can achieve an average of \optSpeedup X speedup on \spec~and an average of \appSpeedup X on real-world applications.

%-------------------------------------------------------------------------------
\section{Acknowledgements}
%-------------------------------------------------------------------------------

We appreciate the anonymous reviewers for their valuable feedback and comments.
This work is supported by the National Natural Science Foundation of China (No. 62141211).

\bibliographystyle{IEEEtran}
\IEEEtriggeratref{30}
\bibliography{\jobname}

\end{document}